\documentclass[lettersize,journal]{IEEEtran}
\usepackage{graphicx}
\usepackage{subcaption}
\usepackage{multirow}
\usepackage{color,soul}
\usepackage[table,xcdraw]{xcolor}
\usepackage{array} % 在导言区加一次就够

\usepackage{CJKutf8}
\usepackage{amsmath} % 支持数学公式
\usepackage{amssymb} % 提供数学符号，例如 \mathbb
\usepackage{url,hyperref,cleveref}
\usepackage{threeparttable}
\usepackage{CJKutf8}
\usepackage{booktabs}
\usepackage{array}
\usepackage{makecell}
\usepackage{wasysym}
\usepackage{todonotes}
\usepackage{tikz}
\usetikzlibrary{shadows, arrows.meta, positioning}
\usepackage[edges]{forest}
\usepackage{xcolor}
\usepackage{pifont}

\definecolor{hiddendraw}{RGB}{205, 44, 36}
\newcolumntype{H}{>{\setbox0=\hbox\bgroup}c<{\egroup}@{}}
\usepackage{amsmath,amsfonts}
\usepackage{algorithmic}
\usepackage{algorithm}
\usepackage{array}
\usepackage{textcomp}
\usepackage{stfloats}
\usepackage{url}
\usepackage{verbatim}
\usepackage{cite}
\begin{document}

\title{A Survey on Speech Large Language Models for Understanding}

% \author{\name{Jing Peng\textsuperscript{1,2*}, Yucheng Wang\textsuperscript{5*}\thanks{\textsuperscript{*} Equal Contribution},Bohan Li\textsuperscript{1}, Yiwei Guo\textsuperscript{1}, Hankun Wang\textsuperscript{1}, YanGui Fang\textsuperscript{4}, Haoyu Li\textsuperscript{1}, Ke Zhang\textsuperscript{6}, Yu Xi\textsuperscript{1,2}\thanks{This work was conducted during the internship of Jing Peng, Yucheng Wang, and Yangui Fang at AISpeech.}, Xu Li\textsuperscript{3}, Xizhuo Zhang\textsuperscript{1,2}, Shuai Wang\textsuperscript{7}, Kai Yu\textsuperscript{1,2\textdagger\thanks{\textdagger Kai Yu is the corresponding author.}}}
% \\
% \address{\textsuperscript{1}
% X-LANCE Lab, School of Computer Science, MoE Key Lab of Artificial Intelligence\\
% Shanghai Jiao Tong University, Shanghai, China\\
% \textsuperscript{2}Jiangsu Key Lab of Language Computing, Suzhou, China\\
% \textsuperscript{3}AISpeech Co., Ltd., Suzhou, China \\
% \textsuperscript{4}Huazhong University of Science and Technology, Wuhan, China \\
% \textsuperscript{5}ETH Zürich, Zürich, Switzerland\\
% \textsuperscript{6}School of Data Science, The Chinese University of Hong Kong, Shenzhen, China\\
% \textsuperscript{7}School of Intelligence Science and Technology, Nanjing University, Suzhou, China}} %\\

\author{Jing Peng\textsuperscript{1,2*}, Yucheng Wang\textsuperscript{3*}\thanks{\textsuperscript{*} Equal Contribution}, Bohan Li\textsuperscript{1,2}, Yiwei Guo\textsuperscript{1,2}, Hankun Wang\textsuperscript{1,2}, YanGui Fang\textsuperscript{5}, Yu Xi\textsuperscript{1,2}, Haoyu Li\textsuperscript{1,2} \thanks{This work was conducted during the internship of Jing Peng, Yucheng Wang, and Yangui Fang at AISpeech.}, Xu Li\textsuperscript{4}, Ke Zhang\textsuperscript{6}, Shuai Wang\textsuperscript{7}, Kai Yu\textsuperscript{1,2\textdagger\thanks{\textdagger Kai Yu is the corresponding author.}}
% Temporary ranking!
\\
\textsuperscript{1}
X-LANCE Lab, School of Computer Science, MoE Key Lab of Artificial Intelligence\\
Shanghai Jiao Tong University, Shanghai, China\\
\textsuperscript{2}Jiangsu Key Lab of Language Computing, Suzhou, China\\
\textsuperscript{3}ETH Zürich, Zürich, Switzerland\\
\textsuperscript{4}AISpeech Co., Ltd., Suzhou, China \\
\textsuperscript{5}Huazhong University of Science and Technology, Wuhan, China \\
\textsuperscript{6}School of Data Science, The Chinese University of Hong Kong, Shenzhen, China\\
\textsuperscript{7}School of Intelligence Science and Technology, Nanjing University, Suzhou, China} %\\

\maketitle

\begin{abstract}
Speech understanding is essential for interpreting the diverse forms of information embedded in spoken language, including linguistic, paralinguistic, and non-linguistic cues that are vital for effective human-computer interaction. The rapid advancement of large language models (LLMs) has catalyzed the emergence of Speech Large Language Models (Speech LLMs), which marks a transformative shift toward general-purpose speech understanding systems. To further clarify and systematically delineate task objectives, in this paper, we formally define the concept of speech understanding and introduce a structured taxonomy encompassing its informational, functional, and format dimensions. Within this scope of definition, we present a comprehensive review of current Speech LLMs, analyzing their architectures through a three-stage abstraction: Modality Feature Extraction, Modality Information Fusion, and LLM Inference. In addition, we examine training strategies, discuss representative datasets, and review evaluation methodologies adopted in the field. Based on empirical analyses and experimental evidence, we identify two key challenges currently facing Speech LLMs—instruction sensitivity and degradation in semantic reasoning—and propose concrete directions for addressing these issues. Through this systematic and detailed survey, we aim to offer a foundational reference for researchers and practitioners working toward more robust, generalizable, and human-aligned Speech LLMs.

\end{abstract}

\section{Introduction}
\label{sec:intro}

\IEEEPARstart{S}{peech} contains a wealth of useful information, ranging from linguistic content, such as lexical meaning and syntax, to paralinguistic cues like emotion and speaking style, as well as non-linguistic context, such as background noise and spatial cues. Understanding speech is therefore an essential task, enabling effective human-machine communication and driving advancements in diverse fields such as virtual assistants, automatic transcription, sentiment analysis, and conversational AI \cite{coucke2018snips, sarikaya2016overview}. Research on speech understanding has a long and evolving history, deeply rooted in the intersection of speech processing and natural language processing (NLP). Its development has spanned decades, transitioning from early rule-based and statistical approaches to modern deep learning and large language model (LLM) paradigms\cite{tur2011spoken}. 

% \subsection{Definition of Speech Understanding}
Although the term “speech understanding” is widely used, there is currently no universally accepted definition. In this survey, we adopt a broad, task-agnostic definition:  
speech understanding can be defined as the integrated process that involves the perception and cognition of spoken language, ultimately producing a textual interpretation. Unlike traditional natural language understanding (NLU), which operates solely on symbolic textual input, speech understanding involves the multimodal interpretation of acoustic signals—including lexical content, prosody, speaker traits, and environmental context. It requires not only recognizing what is said but also perceiving how it is said, who is speaking, and under what conditions. This process entails extracting diverse information types from speech, aligning them with cognitive goals, and producing structured or generative outputs, thereby bridging low-level signal processing with high-level language understanding.

This contrasts with spoken language understanding (SLU), which focuses more narrowly on extracting structured semantics from transcribed speech~\cite{tur2011slu}. In comparison, our definition reflects a more holistic view that treats speech as a multimodal signal for broad understanding and reasoning tasks. In this introduction, we will start with the historical evolution of speech understanding approaches, and subsequently introduce the primary focus of this survey: Speech Large Language Models (Speech LLMs) for understanding.

% \subsection{Development of Speech Understanding Approaches}

Speech understanding was historically approached through a cascaded architecture, where individual subtasks—such as transcribing speech into text via automatic speech recognition (ASR) and extracting semantic information like intent or entities—were handled by separate modules in a sequential pipeline. This design became the standard paradigm from the 1990s through the 2010s, forming the backbone of early research and commercial systems alike~\cite{tur2011spoken, lugosch2019speech}. Over time, ASR and downstream semantic tasks evolved largely independently. ASR research focused on reducing word error rates through advances in acoustic modeling, language modeling, and, later, deep learning-based end-to-end transcription models~\cite{hinton2012deep,graves2014towards,chan2015listenattendspell}. Meanwhile, semantic understanding tasks—often grouped under the term SLU—progressed from rule-based systems to statistical and neural approaches for intent classification and slot filling, frequently borrowing methods from text-based NLP \cite{qin2021survey}. Although researchers recognized the limitations of this modular setup, efforts to optimize the subtasks jointly remained limited. Early approaches included passing ASR confidence scores or N-best hypotheses to the semantic module \cite{yang2015using, tur2013asr, ladhak2016lattice}, while later work explored joint modeling via feature sharing and multi-task learning \cite{haghani2018audio, kim2017joint, serdyuk2018towards}. However, these integrations were generally shallow and constrained by the rigid pipeline structure, which remained dominant throughout this period.

While cascaded architectures have been widely adopted due to their modularity and flexibility, researchers have long identified several fundamental drawbacks. These include the propagation of recognition errors to the understanding component~\cite{lugosch2019speech}, 
increased latency from sequential processing~\cite{arora2022two}, 
loss of valuable acoustic and prosodic information during transcription~\cite{chuang2019speechbert}, 
and the lack of joint optimization across modules~\cite{lugosch2019speech}.
These limitations have motivated a shift toward more unified modeling approaches. In recent years, this has led to the emergence of end-to-end speech language models that integrate recognition and understanding into a single system.

The first major stage in the development of unified speech language models emerged in the late 2010s, characterized by end-to-end systems that map speech directly to high-level semantic representations, such as textual labels, summaries, or task-specific outputs~\cite{serdyuk2018towards, lugosch2019speech, shon2022slue, seo2022end}. These models typically employed either encoder architectures trained from scratch using ASR modeling frameworks, such as CTC, RNN-T, or attention-based encoder-decoder models~\cite{graves2006ctc,graves2012rnnt,amodei2016deep,chan2015listenattendspell}, or self-supervised pretrained encoders such as wav2vec 2.0, HuBERT, and WavLM~\cite{baevski2020wav2vec,hsu2021hubert,chen2022wavlm}. In these cases, encoders are paired with lightweight decoders to produce task-specific semantic outputs. Compared to cascaded pipelines, these models offer reduced error propagation, improved robustness, and greater architectural simplicity. \textbf{However, although task-specific models have achieved relatively mature performance in individual speech understanding tasks, they remain inadequate for the broader challenge of general speech understanding, which requires the ability to process and integrate diverse types of information from speech}. The capacity for complex reasoning and task adaptation of models at this stage remains limited, primarily due to the simple decoders and the lack of general linguistic knowledge. These limitations motivate the next-stage LLM-driven approaches.

% These models offer clear advantages over cascaded pipelines, including reduced error propagation, improved robustness, and greater architectural simplicity. However, they are limited in their ability to perform complex reasoning or adapt to new tasks, due in part to shallow decoders and the absence of general-purpose linguistic knowledge—shortcomings that paved the way for the next stage of development driven by large language models.

Speech LLMs transition speech understanding from the encoder-decoder architectures to language-model-centric frameworks. Leveraging powerful pretrained text-based LLMs, Speech LLMs map speech inputs, which are typically processed by self-supervised encoders, into the textual embedding space. This enables language models to perform reasoning, text generation, and understanding directly from speech. This architecture enables a wide range of downstream tasks, including summarization, question answering, and emotion detection through flexible prompting and instruction tuning. Representative models such as SALMONN~\cite{tang2023salmonn}, SEED~\cite{bai2024seed}, Listen-Think-Understand~\cite{gong2024listenthinkunderstand}, and DESTA-2~\cite{lu2025desta2developinginstructionfollowingspeech} exemplify this paradigm~\cite{ma2024, xu2025fireredasropensourceindustrialgrademandarin, deshmukh2023pengi, geng2025osumadvancingopenspeech}. 
More importantly, Speech LLMs integrate transcription, semantic interpretation, and response generation into a unified framework, eliminating the barrier between ASR and SLU and enabling holistic speech understanding. A detailed discussion of general speech understanding is provided in Section~\ref{sec:evolution}. 

Despite the rapid progress of LLM-centric frameworks, the definition of Speech LLMs remains non-standardized in current research. Although interpretations and research scopes vary, this survey defines Speech LLMs as understanding-oriented spoken language processing. Specifically, we focus on Speech LLMs that generate textual outputs, such as transcriptions, intent labels, summaries, or answers. Alternative definitions may include speech-to-speech generation tasks, which are beyond the scope of this work and are reviewed in other surveys~\cite{arora2025landscapespokenlanguagemodels, ji2024wavchatsurveyspokendialogue}. While several recent surveys have explored the related concepts~\cite{cui2024recent, arora2025landscapespokenlanguagemodels}, none have provided a focused and systematic review from the perspective of speech understanding. \textbf{Beyond serving as a core module in most mainstream Speech LLMs, speech understanding itself has emerged as a distinct research direction, with dedicated Speech LLMs now being developed solely for understanding tasks—a trend gaining significant traction in both academia and industry\cite{deshmukh2023pengi,gong2024listenthinkunderstand,tang2023salmonn,chu2023qwen,chu2024qwen2,lu24c_interspeech}.} Existing surveys primarily focus on speech generation or broad multimodal frameworks, offering limited context for linking Speech LLMs to traditional speech understanding tasks. This survey fills this gap by providing a focused, structured overview of the design, training, and evaluation of Speech LLMs for understanding-oriented tasks.

Our contributions are as follows:
\begin{itemize}
% \item We present the first comprehensive survey on Speech LLMs from the perspective of speech understanding, distinguishing this focus from prior work centered on speech generation or general-purpose multimodal models.
% \item We define and formalize the concept of speech understanding, establishing a foundation that encompasses informational, functional, and format dimensions.
% \item We trace the historical evolution of modeling approaches—covering cascaded pipelines, End-to-End (E2E) specialized speech understanding system, and the rise of LLM-centric general speech understanding systems—and clarify the design space of current systems.
% \item We provide a structured synthesis of model structures, training methodologies, and datasets, alongside a summary of current benchmarks and empirical findings.
% \item We discuss key open challenges based on extensive experiment results, including the instruction sensitivity and degradation of semantic reasoning ability of current Speech LLMs.
\item We present the first comprehensive survey of Speech Large Language Models (Speech LLMs) from the perspective of speech understanding, providing the first systematic conceptualization and task-oriented definition of speech understanding itself, along with a taxonomy spanning informational, functional, and format dimensions.
\item We review the evolution of modeling approaches—ranging from cascaded pipelines to LLM-centric architectures—and provide a structured synthesis of current model designs, training strategies, and datasets, along with an innovative evaluation of their generalization performance across speech understanding tasks.
\item Based on our own experimental analysis, we identify key challenges facing current Speech LLMs—including instruction sensitivity and the first articulation of limitations in semantic reasoning—and outline directions for future research toward more generalizable and robust speech understanding.
\end{itemize}
By centering the discussion on understanding-oriented tasks and offering both conceptual clarity and technical depth, this work serves as a foundational reference that helps researchers in speech processing, spoken language understanding, multimodal modeling, and conversational AI better understand the current landscape and situate new models within it.

% \subsection{Organization of this Survey}
The remainder of this survey is organized as follows. Section~\ref{sec:taxonomy} introduces a taxonomy of speech understanding tasks, defining the scope of speech understanding and categorizing the major task types. Section~\ref{sec:evolution} traces the evolution of Speech LLMs, covering the shift from cascaded pipelines to unified models and contrasting architectural paradigms based on discrete versus continuous speech modeling. Section~\ref{sec:structure} examines the model structure of Speech LLMs, breaking it down into three stages: modality feature extraction, modality information fusion, and LLM inference. Section~\ref{sec:train} reviews the training strategies of current Speech LLMs, while Section~\ref{sec: Datasets} summarizes the key datasets used for pretraining and evaluation. Section~\ref{sec: Performance} discusses performance benchmarks and empirical results across various understanding tasks. Finally, Section~\ref{sec:challenge} outlines open challenges and future directions for advancing deep speech understanding with LLMs. The overall organization is illustrated in Figure~\ref {fig:organization}.

\tikzstyle{leaf}=[mybox,minimum height=3em,
fill=hidden-orange!50, text width=5em,  text=black,align=left,font=\footnotesize,
inner xsep=4pt,
inner ysep=1pt,
]

\begin{figure*}[thp]
  \centering
  \begin{forest}
    forked edges,
    for tree={
      grow=east,
      reversed=true,  % increase counter-clockwise
      anchor=base west,
      parent anchor=east,
      child anchor=west,
      base=left,
      font=\normalsize,
      rectangle,
      draw=hiddendraw,
      rounded corners,
      align=left,
      minimum width=2.5em,
      inner xsep=4pt,
      inner ysep=2pt,
    },
    where level=0{align=center,font=\small}{},
    where level=1{text width=12.8em,align=center,font=\footnotesize}{},
    where level=2{text width=20.5em,align=left,font=\footnotesize}{},
[\textbf{Speech LLMs for Understanding}
  [Sec. \ref{sec:taxonomy}: Taxonomy
    [Sec. \ref{taxonomy:information}: Informational Dimension]
    [Sec. \ref{taxonomy:functional}: Functional Dimension]
    [Sec. \ref{taxonomy:format}: Format Dimension]
  ]
  [Sec. \ref{sec:evolution}: Evolution
    [Sec. \ref{sec: evolution_A}: Emergence of Speech LLM]
    [Sec. \ref{sec: evolution_B}: Structure Evolution of Speech LLM]
  ]
  [Sec. \ref{sec:structure}: Model Structure
    [Sec. \ref{sec:feature}: Modality Feature Extraction]
    [Sec. \ref{sec:fusion}: Modality Information Fusion]
    [Sec. \ref{sec:LLM_inference}: LLM Inference]
  ]
  [Sec. \ref{sec:train}: Training Strategies
    [Sec. \ref{sec:alignment}: Modality Alignment]
    [Sec. \ref{sec:multitask}: Multitask Training]
    [Sec. \ref{sec:preference_alignment}: Preference Alignment]
    [Sec. \ref{sec:discrete_training}: Training of Speech LLMs with Discrete Tokens]
  ]
  [Sec. \ref{sec: Datasets}: Datasets
    [Sec. \ref{sec:traditional_dataset}: Datasets for Perception \& Shallow Cognition Tasks]
    [Sec. \ref{sec:deep_datasets}: Datasets for Deep Cognition Tasks]
    [Sec. \ref{sec:recent_datasets}: Recent Developments in Datasets]
  ]
  [Sec. \ref{sec: Performance}: Performance in SU Tasks
    [Sec. \ref{sec:taxonomy_evaluation}: Taxonomy of Evaluation Methods]
    [Sec. \ref{sec:performance_LLM}: Performance of Speech LLMs Across Tasks]
    [Sec. \ref{sec:real_world_performance}: Model Performance in Real-World Scenarios]
  ]
  [Sec. \ref{sec:challenge}: Challenges
    [Sec. \ref{sec:instr_sensitivity}: Instruction Sensitivity]
    [Sec. \ref{sec:challenge2}: Degradation on Semantic Understanding Ability]
  ]
  [Sec. \ref{sec:print}: Future Exploration]
  [Sec. \ref{sec:conclusion}: Conclusion]
]
\end{forest}
\caption{The organization of this paper is illustrated here. For ease of navigation, this figure offers a concise outline of the paper structure. Readers can refer to the respective sections for full details.}
\label{fig:organization}
\end{figure*}

\section{A Taxonomy of Speech Understanding Tasks}
\label{sec:taxonomy}

Speech understanding refers to the multimodal cognitive process of extracting and interpreting meaningful information from spoken input. It differs from Natural Language Understanding (NLU), which primarily focuses on linguistic cognition by inferring semantic, logical, and affective meaning from symbolic text. 
In contrast, speech understanding integrates a critical perceptual component. The spoken modality carries not only linguistic content but also paralinguistic and non-linguistic information. As such, speech understanding requires models to perceive and interpret rich, temporally dynamic acoustic signals before any semantic reasoning can occur.
\label{sec:definition}
This makes speech understanding a joint endeavor of perception and cognition. Perception provides the raw, multi-dimensional signal basis, while cognition formulates the goals and strategies for interpretation. Crucially, the two are interdependent: effective cognition hinges on accurate, fine-grained perception, and meaningful perception must be guided by cognitive priors and task objectives. This relationship underpins the design of speech understanding systems.

To further structure the concept of speech understanding, we categorize speech understanding tasks from a system-oriented perspective, comprising three complementary dimensions. The \textit{informational} dimension focuses on the types of information embedded in speech input, forming the perceptual front end. The \textit{functional} dimension considers the understanding objectives, defining what cognition tasks the system is expected to perform. Finally, the \textit{format} dimension addresses how these tasks are instantiated in terms of input–output structure, capturing the overall formulation and operationalization of speech understanding systems.

\subsection{Informational Dimension: Categorization by Speech Information Types}
\label{taxonomy:information}
Speech signals convey a rich set of information types, which can be grouped into three broad categories. Based on this dimension, speech understanding tasks can be further categorized according to the primary type of information they aim to interpret.

\begin{itemize}
    \item \textbf{Linguistic Information}: Focuses on transcribing and understanding lexical and syntactic content from speech. Key tasks include Automatic Speech Recognition (ASR) and Spoken Language Translation (SLT), where challenges arise from homophones, dialectal variations, and transcription errors.
    
    \item \textbf{Paralinguistic Information}: Concerned with extracting cues beyond words, such as emotion, prosody, and speaker intent. Related tasks include Emotion Recognition, Speaker Diarization, and Dialogue Act Classification, which rely on features like intonation and speaking style.
    
    \item \textbf{Non-Linguistic Information}: Encompasses contextual acoustic cues supporting interaction analysis. This includes tasks like Voice Activity Detection (VAD), Sound Event Classification, and Acoustic Scene Understanding, which depend on background noise, speaker turns, and spatial cues.
\end{itemize}

This dimension informs the design of front-end speech encoders by highlighting the diverse information types embedded in speech. By distinguishing between linguistic, paralinguistic, and non-linguistic content, it enables targeted modeling strategies for feature extraction, representation learning, and input conditioning in speech understanding systems.

\begin{figure*}[tbhp] % 使用 figure 环境来放置图片
    \centering
    \includegraphics[width=0.95\textwidth]{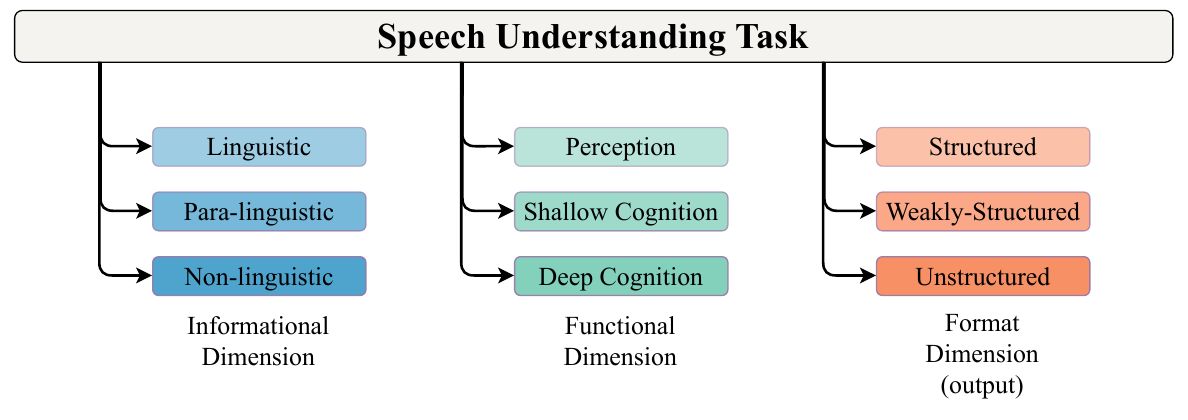} % 插入图片，指定宽度为页面宽度的一半
    \caption{A Three-Dimensional Taxonomy of Speech Understanding Tasks.}
    \label{fig:taxonomy} % 可选：用于引用此图片
\end{figure*}

\subsection{Functional Dimension: Categorization by Cognitive Objectives}
\label{taxonomy:functional}

Speech understanding tasks can be broadly categorized according to their cognitive goals into three progressive levels:

\begin{itemize}
    \item \textbf{Perception Tasks}: Focus on extracting and manipulating audio signals, requiring minimal semantic reasoning. Representative tasks include automatic speech recognition (ASR), keyword spotting (KWS), speaker diarization (SD), and voice activity detection (VAD).
    
    \item \textbf{Shallow Cognition Tasks}: Rely on intuitive, pattern-based understanding, corresponding to fast and heuristic reasoning. Examples include speech translation (ST), emotion classification, intent detection, and sentiment recognition.
    
    \item \textbf{Deep Cognition Tasks}: Require complex, context-aware reasoning, often involving discourse modeling and inference. Typical tasks include complex spoken question answering, conversational summarization, speaker intent explanation in ambiguous contexts, and multimodal reasoning grounded in acoustic and textual context.
\end{itemize}

Our definition of the distinction between shallow and deep cognition tasks is grounded in the \textit{Dual Process Theory of cognition}, which posits two complementary modes of reasoning: System 1, which operates quickly, automatically, and intuitively, and System 2, which engages in slower, more deliberate, and analytical thinking \cite{evans2008dual, kahneman2011thinking}. In this framework, shallow cognition tasks in speech understanding reflect System 1 processes, as they often rely on heuristic or pattern-based interpretation without the need for explicit multi-step reasoning. In contrast, deep cognition tasks correspond to System 2, requiring models to perform structured inference, integrate complex contextual cues, and resolve ambiguity. Framing speech understanding in this dual-process perspective helps provide a principled foundation for stratifying task complexity and guiding model design. By clarifying the cognitive depth required by each task type—from perceptual to deep inference—this dimension facilitates strategic planning of training objectives, curriculum learning stages, and evaluation protocols across varying levels of semantic complexity.

\subsection{Format Dimension: Categorization by Input–Output Structures}
\label{taxonomy:format}

Unlike the information and functional dimensions, which describe what knowledge is used and why a task is performed, the format dimension specifies how inputs are packaged and constrained. Because structure is observable and easy to label, this dimension offers a crisp taxonomy without diluting the main theme. It also aligns with our evaluation suite in Section~\ref{sec:taxonomy_evaluation}.

\subsubsection{Input Formulation}
We can distinguish speech understanding tasks based on the structure of the input conditions. While the core inputs always include the speech signal and natural language instruction, models may also be conditioned on \textit{external information sources} that enhance performance or enable new capabilities. Based on the degree of structure and prior knowledge encoded in these inputs, we categorize them into:

\begin{itemize}
    \item \textbf{Structured Input Sources}: Provide explicitly formatted signals that provide direct guidance to the model. Typical examples include:
    \begin{itemize}
        \item \textit{Hotword lists} (e.g., user-defined keywords for biasing ASR),
        \item \textit{Speaker embeddings} (e.g., fixed-dimensional vectors encoding speaker identity),
        \item \textit{Lexicons or domain constraints} (e.g., task-specific vocabularies or grammar rules),
        \item \textit{Knowledge structures} (e.g., ontologies, documents, or database schemas that encode external domain knowledge in structured form)
    \end{itemize}

    \item \textbf{Unstructured Input Sources}: Provide open-form, contextually derived signals that are not explicitly constrained in format. These may include:
    \begin{itemize}
        \item \textit{Dialogue history} (e.g., preceding conversational turns),
        \item \textit{Background descriptions} (e.g., task instructions, user profiles, or scene summaries), and
        \item \textit{Long-form contextual summaries} (e.g., prior utterance content or semantic memory).
    \end{itemize}
\end{itemize}

\subsubsection{Output Formulation} We can also categorize tasks based on the structure of their output, distinguishing among the following types:
\label{taxonomy:output}
\begin{itemize}
    \item \textbf{Structured Tasks}: Defined by structured outputs that can be rigorously evaluated using formal methods, enabling the computation of precise and objective performance metrics. Common examples include:
    \begin{itemize}
        \item \textit{Automatic speech recognition (ASR)} (e.g., transcribing speech into verbatim or normalized text),
        \item \textit{Speaker analysis} (e.g., speaker identification, verification, or diarization with predefined labels),
        \item \textit{Emotion classification} (e.g., assigning categorical emotional labels),
        \item \textit{Keyword spotting (KWS)} (e.g., detecting predefined terms),
        \item \textit{Voice activity detection (VAD)} (e.g., binary speech/non-speech labeling), and
        \item \textit{Task-oriented structured tagging} (e.g., predicting dialogue acts, user intents, or domain codes).
    \end{itemize}

    \item \textbf{Unstructured Tasks}: Produce open-ended outputs without standardized answers, resisting formal evaluation and depending on contextual or subjective judgment. Representative tasks include:
    \begin{itemize}
        \item \textit{Speaker description} (e.g., generating natural language descriptions of speaker traits such as age or accent),
        \item \textit{Emotion description} (e.g., expressing nuanced emotional states in free-form text),
        \item \textit{Speech quality assessment} (e.g., evaluating fluency or clarity),
        \item \textit{Conversational style analysis} (e.g., inferring politeness or assertiveness),
        \item \textit{User intent explanation} (e.g., interpreting ambiguous or informat utterances), and
        \item \textit{Free-form speech interpretation} (e.g., summarizing or describing content in open language).
    \end{itemize}

    \item \textbf{Weakly-Structured Tasks}: Lie between structured and unstructured outputs, allowing multiple acceptable answers under soft correctness constraints. Such tasks support approximate objective metrics that enable partial formal evaluation. Typical cases include:
    \begin{itemize}
        \item \textit{Speech translation (ST)} (e.g., converting speech from one language to another with diverse valid outputs, evaluated by BLEU~\cite{papineni-etal-2002-bleu, barrault2023seamlessm4t}),
        \item \textit{Speech summarization} (e.g., generating condensed content where factual coverage and coherence are evaluated using ROUGE~\cite{lin2004rouge}).
    \end{itemize}
\end{itemize}

This framework not only enables a comprehensive categorization of existing tasks but also provides guidance for the design of future speech understanding systems. Structured outputs benefit from precise metrics and strong supervision, while unstructured outputs better reflect real-world complexity and the need for generative reasoning. In particular, structured outputs are directly interpretable by downstream computational systems and allow reliable, objective evaluation, whereas unstructured outputs require generative understanding and are not straightforwardly machine-actionable. Similarly, integrating structured and unstructured input sources enhances model controllability, personalization, and adaptability in diverse environments.

Taken together, when situated in the context of Speech LLMs, this taxonomy not only provides a principled framework for organizing current capabilities, but also highlights the evolving demands of future systems. With increasing emphasis on multi-task generalization, instruction following, and context-aware reasoning, Speech LLMs must flexibly integrate diverse input types and generate outputs of varying structure and specificity. The taxonomy thus informs the development of more adaptive, interpretable, and goal-aligned architectures. It supports a shift from narrow ASR-centric modeling to a broader paradigm of holistic spoken language understanding.

\section{Toward General Speech Understanding: Emergence and Evolution of Speech LLMs}\label{sec:evolution}

To provide a clearer exposition of the development trajectory of Speech LLMs, this section unfolds the discussion along two complementary dimensions. Section~\ref{sec: evolution_A} presents a longitudinal view that traces the historical evolution of speech understanding systems, highlighting the emergence and significance of Speech LLMs in this context. In parallel, Section~\ref{sec: evolution_B} offers a structural perspective, characterizing the progression and iterations of Speech LLMs across two different architectural paradigms.

\subsection{From Cascade to Unified: The Emergence of Speech LLMs for Speech Understanding}
\label{sec: evolution_A}
Following the definition of speech understanding in Section~\ref{sec:definition} and the task taxonomy proposed in Section~\ref{taxonomy:functional}, we can categorize speech understanding into two broad perspectives: one centered on perceptual processing  (e.g., ASR and related tasks), and the other focused on cognitive reasoning over spoken input, encompassing both shallow and deep understanding objectives (e.g., SLU and beyond). Given the absence of a consistent historical term, this section traces the evolution of Speech LLMs from the representative perspectives of ASR and SLU, which jointly illuminate the broader trajectory of speech understanding. While not exhaustive, these two lines of development offer illustrative anchors for understanding the structural and functional progress in the field.

\begin{figure*}[htbp]
    \centering
    \includegraphics[width=1.0\textwidth]{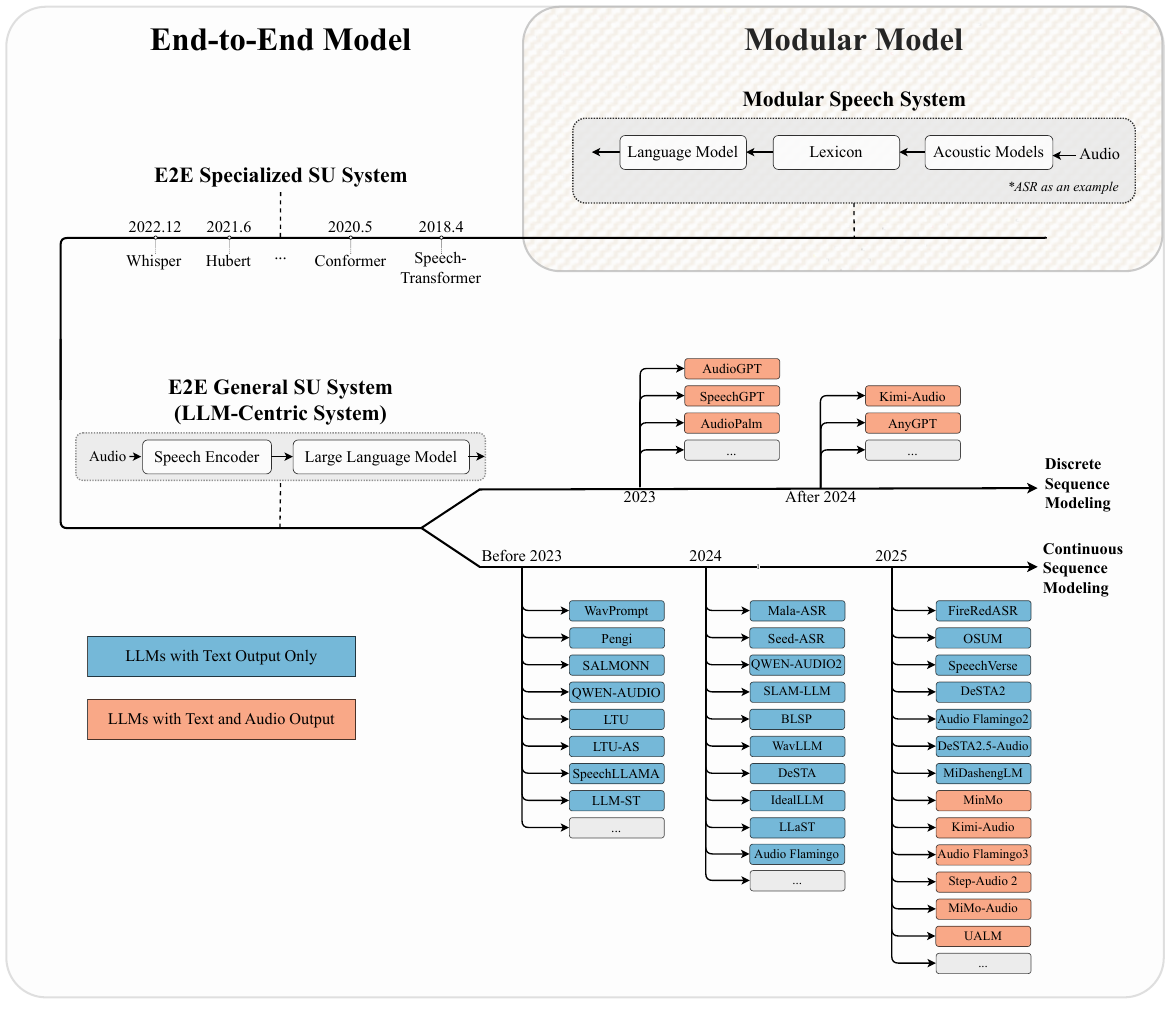} 
    \caption{
The structural evolution of Speech LLMs, illustrating the transition from modular architectures to end-to-end frameworks. Early modular systems are exemplified by traditional ASR pipelines. The end-to-end paradigm subsequently diverges into two branches: E2E Specialized Speech Understanding System and E2E General Speech Understanding System. E2E General Speech Understanding systems, also regarded as \textit{LLM-Centric} systems, are further categorized into discrete sequence modeling (top) and continuous sequence modeling (bottom), reflecting their strategies of representing features. }
    \label{fig:development}
\end{figure*}

We broadly divide the development of models for speech understanding into two main stages. The first is the era of \textit{modular architectures}, typically characterized by cascaded structures with separately optimized components. The second is the emergence of \textit{end-to-end architectures}, where the system is trained holistically to optimize performance across the entire pipeline. Within the end-to-end paradigm, a further transition has occurred. The field has moved from \textit{E2E Specialized Speech Understanding Systems}, which are designed for a single or a small set of well-defined tasks, to \textit{E2E General Speech Understanding Systems}. In these general systems, language instructions can specify any task, and the model supports outputs that can capture any form of understanding. Each stage not only represents a milestone in model capacity and architecture but also demonstrates the increasing convergence between tasks, ultimately culminating in a unified modeling capability aimed at general speech understanding. An overview of this development process is illustrated in Figure~\ref{fig:development}.

\subsubsection{Modular Architecture}

Early efforts in speech processing largely followed a modular architecture. For example, in ASR, the pipeline typically involved a separately trained acoustic model (AM), language model (LM), and decoder, integrated via weighted finite-state transducers (WFSTs). These components were optimized independently and lacked task adaptability. Typical implementations include Gaussian Mixture Model - Hidden Markov Model (GMM-HMM) frameworks and later Deep Neural Network (DNN-HMM) hybrids~\cite{hinton2012deep, dahl2012context}, which formed the foundation of many large-scale ASR toolkits such as Kaldi~\cite{povey2011kaldi} and CMU Sphinx~\cite{lamere2003cmusphinx}. This design required extensive engineering effort, heavily relied on handcrafted configurations, and lacked the capacity to integrate knowledge holistically or support multitask generalization.

Moreover, in the SLU domain, spoken language understanding was conventionally built on top of ASR outputs. Early SLU systems adopted the \textit{ASR $\rightarrow$ NLU} pipeline~\cite{tur2011slu, mesnil2014rnnslu}, where natural language understanding models operated on 1-best or N-best ASR hypotheses to perform tasks such as intent classification, slot filling, and dialogue state tracking~\cite{hakkani2016multi}. However, such cascaded systems suffer from error propagation and degraded performance in noisy or ambiguous speech scenarios. In addition, rich prosodic and acoustic cues were often lost during transcription, limiting deeper semantic interpretation.

Other models for specific speech understanding tasks, such as emotion recognition, have similarly undergone a modular development phase. Although these approaches provide strong interpretability, they still exhibit substantial potential for improvement in performance.

\subsubsection{E2E Specialized Speech Understanding System}
% to do: 定义 ASR+SLU级联，端到端演进，定义清楚specialized。切入点：解决什么任务，其他任务也需要带过，比如emotion，神经网络时代，多种理解任务出现融合趋势，着重看asr和slu的变迁，其他的也相似，音频进入，当前目标结果作为输出。性能不错，但是数据量大，且多任务并不够通用。LLM的specialization。开头结尾衔接清楚。
With the development of deep neural networks, there has been a shift toward end-to-end architectures in models designed for specific speech understanding tasks: from modular pipelines to end-to-end architectures, and from cascaded processing to unified modeling. We refer to this class of models, which focus on solving one or a few specific tasks in an end-to-end manner, as \textit{E2E Specialized Speech Understanding System}.
% These developments reflect a broader trend toward joint optimization and holistic representation learning in speech understanding systems.

The \textit{first major stage} in this transition is the rise of end-to-end ASR architectures. Early E2E systems employed frame-synchronous models based on Connectionist Temporal Classification (CTC)~\cite{graves2006ctc}, which simplified training by removing the need for frame-level alignment~\cite{amodei2016deep}. These were followed by Recurrent Neural Network Transducer (RNN-T)~\cite{graves2012rnnt}, which relaxed independence assumptions and supported streaming applications. In parallel, attention-based encoder-decoder (AED) models such as Listen, Attend and Spell (LAS)~\cite{chan2015listenattendspell}, SpeechT5~\cite{ao2022speecht5}, and Whisper~\cite{radford2023robust} introduced autoregressive decoding and enabled multitask learning capabilities. These systems marked a substantial improvement over traditional pipelines by integrating acoustic and linguistic modeling into a single trainable framework.

Complementing these architectural advancements, the E2E era also witnessed significant perceptual progress through self-supervised learning (SSL). Early hand-crafted features such as MFCCs were gradually replaced by rich, contextualized representations learned by models such as wav2vec 2.0~\cite{baevski2020wav2vec}, HuBERT~\cite{hsu2021hubert}, and WavLM~\cite{chen2022wavlm}. These SSL-based encoders substantially enhanced robustness and generalization and became foundational in many state-of-the-art E2E systems~\cite{yang2021superb}. However, despite these improvements, conventional E2E models often rely on relatively shallow decoders with limited linguistic capacity, making them less suitable for tasks that require long-range reasoning or rich semantic interpretation.

In the SLU area, the end-to-end transition involved models that directly predicted semantic structures—such as intents and slots—from raw speech without relying on intermediate transcriptions. Pioneering works like Joint SLU~\cite{lugosch2019speech} and SLU-Transformer~\cite{qian2021sluformer} demonstrated that speech encoders pretrained via SSL could be effectively paired with semantic decoders to form compact, robust pipelines. 

As the capabilities of end-to-end models improved, their target tasks evolved toward more aggregated objectives. While these models enhanced generalization and task-specific performance, they continued to face challenges in data efficiency and transferability across tasks.

\begin{figure*}[t]
    \centering
    \includegraphics[width=1.0\textwidth]{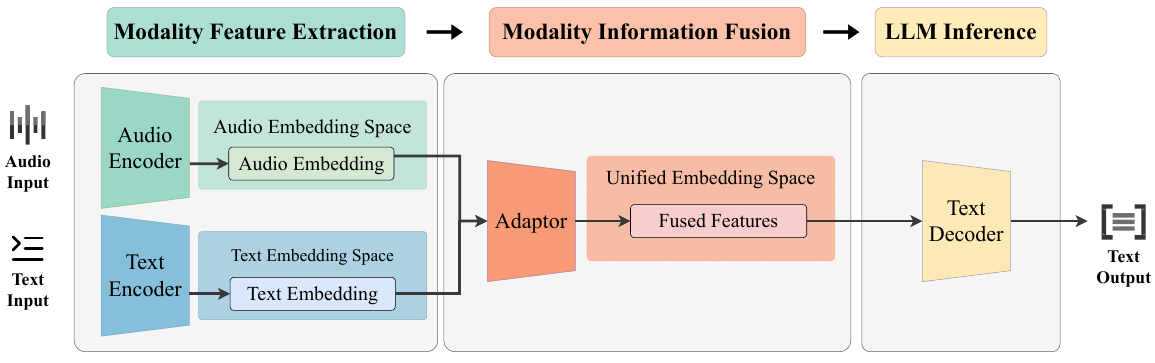} 
    \caption{Overview of Speech LLM Architectures with Speech and Text Inputs and Text Outputs.}
    \label{fig:structure}
\end{figure*}

\subsubsection{E2E General Speech Understanding System}
% 明确描述，界定 general speech understanding的区别
To address the need for richer and more general speech understanding tasks, a more recent trend within the end-to-end paradigm is the development of \textit{E2E General Speech Understanding Systems} or \textit{LLM-Centric Systems}, which integrate LLMs as core reasoning components. Speech understanding models centered around LLMs reflect a paradigm shift toward cognition-guided perception. This trend marks an important development in the evolution of multimodal systems, where the language model no longer holds an equal status with the perceptual front-end but instead serves as the central controller of interpretation. The conceptual roots of this perspective can be traced back to~\cite{liu2020modular}. These systems represent a structural and functional evolution from conventional E2E models, shifting the focus from acoustic-driven generation to language-model-driven interpretation. Unlike traditional AED-based architectures, \textit{LLM-Centric} systems align speech inputs to the text modality and utilize LLMs for decoding, generation, and reasoning~\cite{tang2023salmonn, ma2024, bai2024seed, xu2025fireredasropensourceindustrialgrademandarin}. This cross-modal design not only enables transfer to diverse downstream tasks—such as summarization, translation, and question answering—but also brings enhanced modularity, interpretability, and instruction-following capability~\cite{gong2024listenthinkunderstand, geng2025osumadvancingopenspeech, deshmukh2023pengi, lu2025desta2developinginstructionfollowingspeech}.

Recent advances in \textit{LLM-Centric} architectures have enabled unprecedented levels of task generalization and language comprehension in speech understanding. It is important to distinguish true \textit{LLM-Centric} approaches from conventional cascaded systems that merely connect ASR outputs to an LLM. While such pipeline-based designs~\cite{huang2023spokenCoT, bapna2022slam} allow downstream reasoning over text, they do not fundamentally integrate speech into the LLM framework and thus fall short of true \textit{LLM-Centric} modeling. In contrast, recent instruction-tuned models such as Listen-Think-Understand~\cite{gong2024listenthinkunderstand} and SALMONN~\cite{tang2023salmonn} exemplify \textit{LLM-Centric} design: they take raw speech as input and directly generate semantically structured outputs for tasks such as intent inference, emotion detection, summarization, and knowledge-grounded question answering. These systems align speech representations with the powerful textual priors embedded in LLMs, enabling prompting, interpretation, and response generation within a unified representational space. This paradigm breaks down traditional barriers within speech understanding tasks, marking a shift toward truly holistic and integrated speech understanding systems.

The evolution from modular pipelines to end-to-end systems, and further into \textit{LLM-Centric} frameworks, charts a trajectory of increasing unification across a wide range of speech understanding tasks like ASR and SLU. While ASR emphasizes accurate linguistic transduction and SLU targets semantic reasoning, both perspectives have progressively converged under the umbrella of Speech LLMs. These models offer a shared representational space and a unified decoding mechanism, making it possible to perform transcription, understanding, and generation in a cohesive framework. In this light, the emergence of \textit{LLM-Centric} Speech LLMs marks a pivotal point in the development of general-purpose spoken language understanding—a capability referred to as \textbf{Speech Understanding (SU)} in this paper, as illustrated in Section ~\ref{sec:definition} in detail.

\subsection{ Structural Evolution of Speech LLM: Discrete vs. Continuous Modeling} 
\label{sec: evolution_B}

A key distinction in the development of Speech LLMs lies in how the speech and text modalities are integrated. This paper categorizes current approaches into two primary paradigms based on the representation format used to bridge audio and language: one relies on discrete tokenization of speech, inspired by the token-based processing paradigm of LLMs in text; the other maintains the speech modality in its continuous embedding form. These two paradigms are orthogonal in nature and reflect fundamentally different strategies for cross-modal alignment. (see Figure~\ref{fig:overall}). 

The first, \textit{discrete sequence modeling}, represents a paradigm in which continuous speech signals are transformed into sequences of quantized units—discrete acoustic tokens that resemble textual tokens in format and function. These discrete tokens are typically obtained through unsupervised or self-supervised speech representation models followed by vector quantization mechanisms. Popular approaches include HuBERT~\cite{hsu2021hubert}, w2v-BERT~\cite{chung2021w2v}, EnCodec~\cite{defossez2022high}, and wav2vec-U~\cite{baevski2021unsupervised}, which learn to capture salient acoustic patterns and discretize them into a token vocabulary.

By converting speech into tokenized sequences, these models enable LLMs to process speech data using the same autoregressive or seq2seq frameworks designed for text. The resulting discrete units can be aligned with textual tokens or used directly as input/output for generation tasks. This token-level compatibility bridges the modality gap and facilitates seamless speech-text integration.

 Representative studies that model speech purely via discretized acoustic tokens include SpeechGPT and AudioGPT~\cite{zhang2023speechgpt, huang2024audiogpt}. These methods quantize continuous waveforms into discrete code sequences (e.g., with neural codecs) and train a generative language model over the token streams, enabling speech understanding and synthesis within a unified, text-like modeling framework.

This modeling approach offers several advantages. It allows speech processing to benefit directly from LLM pretraining by transforming the problem into a format that LLMs are optimized for: token sequences. Moreover, it enables understanding and generation within the same unified framework. However, models that rely solely on discrete token representations to encode speech information tend to exhibit degraded perceptual quality in speech perception.
First, their performance on semantic information perception tasks such as ASR is often inferior to that of traditional models like Whisper or HuBERT. Second, these models also show limited capability in perceiving fine-grained paralinguistic cues and other subtle aspects of speech. Building upon these considerations, this work does not primarily focus on purely discrete sequence modeling approaches.
% Given that this paper centers on speech understanding rather than generation, we highlight these constraints but do not elaborate on generation-specific issues.

The second paradigm, \textit{continuous sequence modeling}, takes a fundamentally different approach by maintaining the speech signal in its continuous embedding form throughout the interaction with the language model. Instead of discretizing the audio, it extracts high-dimensional features from pretrained encoders (e.g., WavLM~\cite{chen2022wavlm}, Whisper~\cite{radford2023robust}, or Conformer~\cite{gulati2020conformer}) and directly feeds them into the LLM, typically through a learned projection or lightweight adaptation layer.

This strategy preserves the rich temporal and spectral information inherent in speech and enables a smoother alignment with the LLM's input space. Notably, \textit{Pengi}~\cite{deshmukh2023pengi} demonstrates how frozen LLMs can be prompted using projected speech embeddings to perform a variety of tasks. Subsequent works like \textit{SALMONN}~\cite{tang2023salmonn}, \textit{Qwen-Audio}~\cite{chu2023qwen}, and \textit{OSUM}~\cite{geng2025osumadvancingopenspeech} further extend this design to support broad speech understanding and multimodal reasoning tasks. 

Continuous modeling has proven particularly effective for applications requiring fine-grained acoustic discrimination, including ASR~\cite{xu2025fireredasropensourceindustrialgrademandarin}, ST~\cite{chen2024llastimprovedendtoendspeech}. Models like \textit{FireRedASR}~\cite{xu2025fireredasropensourceindustrialgrademandarin} and \textit{LLAST}~\cite{chen2024llastimprovedendtoendspeech} show that state-of-the-art performance can be achieved with minimal architectural modification. In many cases, a simple projection layer is sufficient, together with limited fine-tuning~\cite{ma2024, bai2024seed}.

Although this paradigm can achieve high accuracy in specific tasks, it also introduces certain challenges. When the amount of training data or task diversity is limited, the model is prone to overfitting~\cite{fang2025lowresourcedomainadaptationspeech, 11010998}. Moreover, when integrating with generative modules, it often requires stronger data-driven support or assistance from discrete modeling approaches ~\cite{chen2025minmo, wu2025step}.
Given its scalability and its alignment with the trend of reusing pretrained language models~\cite{fathullah2024prompting, yang2024mala}, continuous modeling has become an increasingly important approach in recent Speech LLM research. This paper, therefore, primarily focuses on continuous sequence modeling, while offering comparative insights into discrete token-based alternatives.

% In the Section \ref{sec: Performance}, we provide a more detailed analysis of representative systems and techniques that have advanced the field across these dimensions.
% 第一个角度：
% 语义理解上，speech llm与llm有没有较为清晰的差异（可以设计实验）。直接跟基座模型来比，或者有没有没有基座或者pretrained好的来看。
% 第二个角度：non-linguistic（区别于pure text）这样的信息的学习或者理解能力怎么样（在这些任务上效果怎么样），与对应的小模型来比较，difference怎么样。与表格相对应，后续进一步量化分析。
% speech encoder能保留多少东西，llm能学到多少。
% 第三个角度：到底该怎么训练一个speech llms（与训练part对应），如何speech-text双模态更好地对应起来，提供一个角度。（有的只做speech，有的只做proj，有的只做llm lora，或者都要，其实公司的ASR的系列实验结果是可以拿来分析的）方法上面整体有没有一个较优路线，是否存在问题。是否有区分方法的结论。【给读者收获】才好进一步说未来的发展方向，有理有据。强调三代架构，要讲清楚，不是局限在asr，asr只是一个例子：
% 异构级联系统
% 一体化端到端（黑箱）
% llm centric跨模态系统->通用任务，所以与纯文本模型可比。利用语言能力结合更丰富语音信息才能更好地完成任务。

% \end{enumerate}
%2025/2/26
\begin{center}
 \begin{table*}[htbp]
\small
\renewcommand{\arraystretch}{1.2}
\centering
\caption{Architectural Summary of Continuous Sequence Modeling Speech LLM Models}
\label{tab:continuous_speech_llms}
\begin{tabular}{l|lll}
\toprule
\textbf{Model} & \textbf{Audio Encoder} & \textbf{Adapter} & \textbf{Decoder} \\
\midrule
\multicolumn{4}{c}{\textbf{Speech LLM with text-only output}} \\
\midrule
WavPrompt~\cite{gao2022wavprompt} & Wav2Vec 2.0 Base Encoder & Linear & GPT-2 (124M) \\
Pengi~\cite{deshmukh2023pengi} & CLAP Audio Transformer & Linear + Gelu + Dropout + Layernorm & GPT-2 (124M) \\
PandaGPT~\cite{su2023pandagpt} & ImageBind Audio Encoder & Linear & Vicuna-7B \\
LTU~\cite{gong2024listenthinkunderstand} & CAV-MAE & Avg-Pool + LayerNorm + Linear & LLaMA-7B \\
SALMONN~\cite{tang2023salmonn} & Whisper-Large-v2 + BEATs Encoder & Q-Former & Vicuna-13B \\
Speech LLaMA~\cite{wu2023decoder} & Light Audio Encoder\textsuperscript{1} & CTC Compression + Linear & LLaMA-7B \\
LTU-AS~\cite{gong_ltuas} & Whisper-Large Encoder + TLTR & Linear & Vicuna-7B \\
LLM-ST~\cite{huang2023speech} & Whisper-Large-v2 Encoder\textsuperscript{2} & -- & GPT-3-13B \\
Qwen-Audio~\cite{chu2023qwen} & Whisper-Large-v2 Encoder & Linear & Qwen-7B \\
SLAM-ASR~\cite{ma2024embarrassingly} & Configurable Encoder\textsuperscript{3} & Configurable Projector\textsuperscript{3} & Configurable LLM\textsuperscript{3} \\
Audio Flamingo~\cite{kong2024audio} & ClapCap Encoder & Gated Cross Attention & OPT-IML-MAX-1.3B \\
BLSP~\cite{wang2023blsp} & Whisper-Large Encoder & 1-D CNN + Linear + Gelu + Linear & LLaMA-2-7B \\
Speechverse~\cite{das2024speechverse} & A Pretrained Audio Encoder\textsuperscript{4} & 1-D CNN & A Pretrained LLM\textsuperscript{4} \\
Mala-ASR~\cite{yang2024mala} & WavLM-Large-v2 Encoder & Linear + ReLU + Linear & LLaMA-based \\
DeSTA~\cite{lu24c_interspeech} & Whisper-Large-v2 Encoder & CNN + Q-Former + Whisper Decoder & LLaMA-2-Chat-7B \\
Qwen2-Audio~\cite{chu2024qwen2} & Whisper-Large-v3 Encoder & Avg-Pool + Linear & Qwen2-7B \\
SeedASR~\cite{bai2024seed} & LUISE Encoder & A Certain Projector\textsuperscript{3} & A Pretrained LLM\textsuperscript{3} \\
E-chat~\cite{xue2024chat} & HuBERT Encoder & Transformer & Baichuan2-7B-Chat \\
LLaST~\cite{chen2024llast} & mHuBERT + Whisper-Large Encoder & MLP & LLaMA2-7B-Instruct \\
WavLLM~\cite{hu2024wavllmrobustadaptivespeech} & Whisper-Large-v2 + WavLM-Large Encoder & 1-D CNN + Bottleneck + Linear & LLaMA-2 \\
IdealLLM~\cite{xue2024ideal} & Whisper-Large-v2 + MMS Encoder & Transformer + CNN + Linear & Phi-3-mini-3.8B \\
DeSTA2~\cite{lu2025desta2developinginstructionfollowingspeech} & Whisper-Large-v2 Encoder & Q-Former + Whisper Decoder & LLaMA-3-8B \\
FireRedASR-LLM~\cite{xu2025fireredasropensourceindustrialgrademandarin} & Conformer Encoder & Linear + ReLU + Linear & Qwen2-Instruct-7B \\
OSUM~\cite{geng2025osumadvancingopenspeech} & Whisper-Medium Encoder & 1-D CNN + Transformer & Qwen2-7B \\
Audio Flamingo 2~\cite{ghosh2025audio} & AF-CLAP Encoder & Gated Cross Attention & Qwen2.5-3B \\
LegoSLM~\cite{ma2025legoslmconnectingllmspeech} & USM Encoder & CTC Layer & Gemma-2B \\
DeSTA2.5-Audio~\cite{lu2025desta2} & Whisper-large-v3 Encoder & Q-Former & Llama-3.1-8B-Instruct\\
MiDashengLM~\cite{dinkel2025midashenglm} & Dasheng Audio Encoder & MLP & Qwen2.5-Omni-7B Thinker\\
\midrule
\multicolumn{4}{c}{\textbf{Speech LLM with text \& speech output}} \\
\midrule
MinMo~\cite{chen2025minmo} & SenseVoice-large Encoder & Transformer + CNN & Qwen2.5-7B-instruct\\
Qwen2.5-Omni~\cite{xu2025qwen2} & Whisper-large-v3–based Encoder & LayerNorm + Avg-Pool + Linear & Qwen2.5-7B\\
Kimi-Audio~\cite{ding2025kimi} & Whisper-Large-v3 EN + GLM4 Tokenizer\textsuperscript{6} & A Certain Downsampler\textsuperscript{5} & 7B Pretrained LLM\textsuperscript{5} \\
Audio Flamingo 3~\cite{goel2025audio} & AF-Whisper Audio Encoder & -- & Qwen-2.5-7B \\
Step-Audio 2~\cite{wu2025step} & Custom Audio Encoder\textsuperscript{7} & Custom Adapter\textsuperscript{7} & Custom LLM\textsuperscript{7} \\
MiMo-Audio~\cite{coreteam2025mimoaudio} & Custom Audio Encoder\textsuperscript{8} & Custom Adapter\textsuperscript{8} & MiMo-7B-Base \\
UALM~\cite{tian2025ualm} & AF-Whisper Audio Encoder & MLP & Qwen2.5-7B \\
Qwen3-Omni~\cite{xu2025qwen3omnitechnicalreport} & AuT\textsuperscript{9} & Custom Adapter\textsuperscript{9} & MoE Transformer\textsuperscript{9} \\
\bottomrule
\end{tabular}

\vspace{0.2cm}
\captionsetup{font=footnotesize}
\caption*{\textsuperscript{1} Light Audio Encoder refers to a lightweight encoder used behind a CTC alignment layer.\\
\textsuperscript{2} Whisper-Large-v2 Encoder without a 30-second padding limit.\\
\textsuperscript{3} Configurable Encoder includes Whisper, WavLM, and HuBERT. Configurable Projector includes Linear + ReLU + Linear, and Q-Former. Configurable LLM includes Vicuna-7B, etc.\\
\textsuperscript{4} Details about specific pretrained models are not publicly disclosed.\\
\textsuperscript{5} Certain components are not detailed in the paper; presumed to be standard linear downsampling or Transformer-based adapters.\\
\textsuperscript{6} The encoder of Kimi-Audio combines discrete semantic tokens and continuous whisper features, making it a hybrid model that does not strictly follow continuous sequence modeling for Speech LLMs. Despite this, we include it in the table for architectural comparison.\\
\textsuperscript{7} Step-Audio 2 uses a frozen pretrained latent audio encoder with a trainable 2x downsampling adaptor, and a custom multimodal LLM decoder that outputs interleaved text and audio tokens.\\
\textsuperscript{8} Uses the MiMo-Audio-Tokenizer’s 32-layer Transformer audio encoder with RVQ-based discretization as the adapter.\\
\textsuperscript{9} AuT refers to a custom Audio Transformer. The adapter is not clearly specified, while the decoder is implemented as a custom MoE Transformer..

For models where specific implementation details are not publicly available through the official GitHub repositories or described in the original papers, the corresponding entries are left unspecified in the table.}
\end{table*}   
\end{center}

\section{Model structure} % Conclude the structure in a neat way and split it up.
\label{sec:structure}

To further understand how Speech LLMs process and reason over spoken input, we examine their underlying architectural design. In particular, we investigate the overall structure of Speech LLMs, which typically follows a modular pipeline comprising distinct stages of processing. Various Speech LLM architectures have been developed up to now, all of which are structured around three fundamental stages: \textbf{Modality Feature Extraction}, \textbf{Modality Information Fusion}, and \textbf{LLM Inference}, as illustrated in~\Cref{fig:structure}.

All Speech LLMs begin by encoding the input information, regardless of the modality. Typically, these Speech LLMs leverage pretrained audio and text encoders to obtain representations of both modalities. The encoded audio and text embeddings are then fused to prepare them for input into the language model. This step transforms the outputs of the feature extraction process into a format compatible with the language model’s input requirements. For Speech LLMs that generates text output, which are also the primary focus of this survey, the LLM inference process typically revolves around a text decoder that receives the fused multimodal features as tokenized input to produce text. For models that optionally support audio output, an LLM decoder is trained to generate audio tokens, which are subsequently converted into speech using a vocoder. However, in this survey, we only focus on the Speech LLMs with text output.

\subsection{Modality Feature Extraction}
\label{sec:feature}
As mentioned in the previous section, during the initial feature extraction stage, the audio is processed in two distinct ways. These approaches primarily differ in the output representation format of the audio features. \textbf{Continuous Sequence Modeling}, which is more commonly adopted in current Speech LLMs, encodes audio using a pretrained and fine-tuned audio encoder, producing an audio embedding $\mathbf{Z}_a\in \mathbb{R}^{B\times T \times D}$ where $B$, $T$, and $D$ represent batch size, number of frames, and embedding dimension. In contrast, \textbf{Discrete Sequence Modeling} discretizes the input audio signal into a sequence of discrete audio tokens—comparable to text tokens derived from raw text—which are not represented in embedding form. In this section, we summarize common models and techniques used to extract audio features using these two approaches.
\subsubsection{Continuous Sequence Modeling}
Continuous sequence modeling is generally more straightforward and typically encodes audio in the form of a time-frequency representation extracted using traditional signal processing-based feature extraction methods from the raw audio signal. Given an audio input $X_a$, a pretrained audio encoder is applied to generate the audio features:
\begin{equation}
    \mathbf{Z}_a = g(\mathbf{X}_a),
\end{equation}
where $g$ denotes the audio encoding process. The most commonly used audio encoders in mainstream Speech LLMs are Whisper~\cite{radford2023robust} and Conformer~\cite{gulati2020conformer}. Whisper is a sequence-to-sequence Transformer model trained on a wide range of speech processing tasks, and its encoder is recognized as one of the most powerful for audio input processing. Conformer is another widely adopted architecture that combines convolutional layers with Transformers to effectively model both local and global dependencies in speech signals, particularly for automatic speech recognition tasks. Additional encoders used in mainstream Speech LLMs include WavLM~\cite{chen2022wavlm}, a predictive, self-supervised learning (SSL) pretrained model. In this stage, aside from the audio encoder itself, subsampling modules are often applied to reduce the temporal resolution of the input features, thereby decreasing computational costs and enabling the model to focus on higher-level representations. Ultimately, the output audio feature embeddings $\mathbf{Z}_a\in \mathbb{R}^{B\times T \times D}$ serve as the input for the next stage. In the context of continuous sequence modeling, Table~\ref{tab:continuous_speech_llms} provides a detailed summary of representative models, organized according to their architectural components.

\subsubsection{Discrete Sequence Modeling}
Discrete sequence modeling converts raw audio into a sequence of discrete tokens that represent acoustic content and can typically be decoded back into high-quality audio. The core idea behind generating discrete tokens lies in vector quantization. Building on VQ-VAE~\cite{van2017neural}, which introduced the concept of encoding continuous audio features into symbolic representations via a learned codebook, current mainstream methods for generating discrete tokens include self-supervised pretrained audio tokenizers such as HuBERT~\cite{hsu2021hubert}, and neural codec models such as EnCodec~\cite{defossez2022high, zeghidour2021soundstream}. A general process involves first encoding raw audio into a sequence of latent representations, as in continuous sequence modeling: $\mathbf{Z}_a = g(\mathbf{X}_a)$, where $\mathbf{Z}_a = [\textbf{z}_1,\textbf{z}_2,\dots,\textbf{z}_T]$ with $\textbf{z}_t\in \mathbb{R}^d$. Next, a quantization function $q(\cdot)$ is applied to obtain the final discrete audio tokens: 
\begin{equation}
    u_t = q(\textbf{z}_t)\ \text{with}\ \textbf{e}_{u_t} = \mathcal{Q}(\textbf{z}_t).
\end{equation} 
Here, $\mathcal{Q}$ denotes the quantization function that maps continuous vectors to codebook vectors $\textbf{e}_{u_t}\in \mathbb{R}^d$, resulting in $u_t$, the discrete token index from the vocabulary set. The final output is a sequence of discrete audio token indices: $U = [u_1, u_2, \dots, u_T]$. Specifically, self-supervised pretrained audio tokenizers, such as HuBERT, learn contextualized audio representations via masked prediction tasks and discretize intermediate features using k-means clustering. The resulting cluster indices serve as phoneme-like discrete tokens. Neural codec models, such as EnCodec, adopt an encoder–quantizer–decoder architecture in which the audio is compressed into discrete tokens using residual vector quantization and later reconstructed~\cite{gray1984vector}. These models are trained end-to-end and produce high-fidelity audio that is well suited for generative tasks. The resulting discrete audio tokens are processed alongside tokenized text, with the integration method detailed in the following section.

\begin{figure*}[htbp]
    \centering
    \begin{subfigure}[t]{\textwidth}
        \centering
        \includegraphics[width=0.85\textwidth]{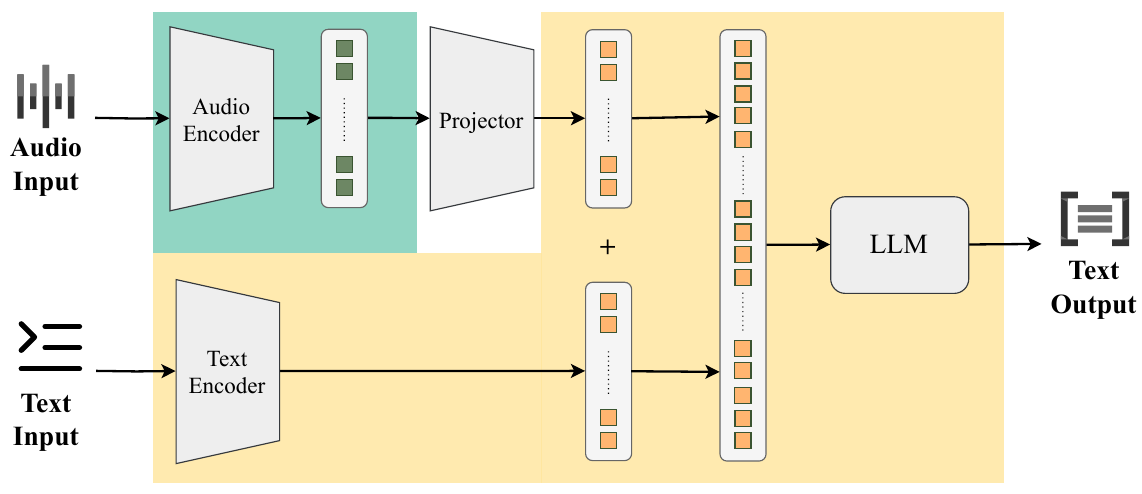}
        \caption{Modality Fusion of Continuous Embedding: Pre-Decoder Alignment}
        \label{fig:direct_projection}
    \end{subfigure}
    
    \vspace{1em} % Adds vertical space between the images
    
    \begin{subfigure}[t]{\textwidth}
        \centering
        \includegraphics[width=0.85\textwidth]{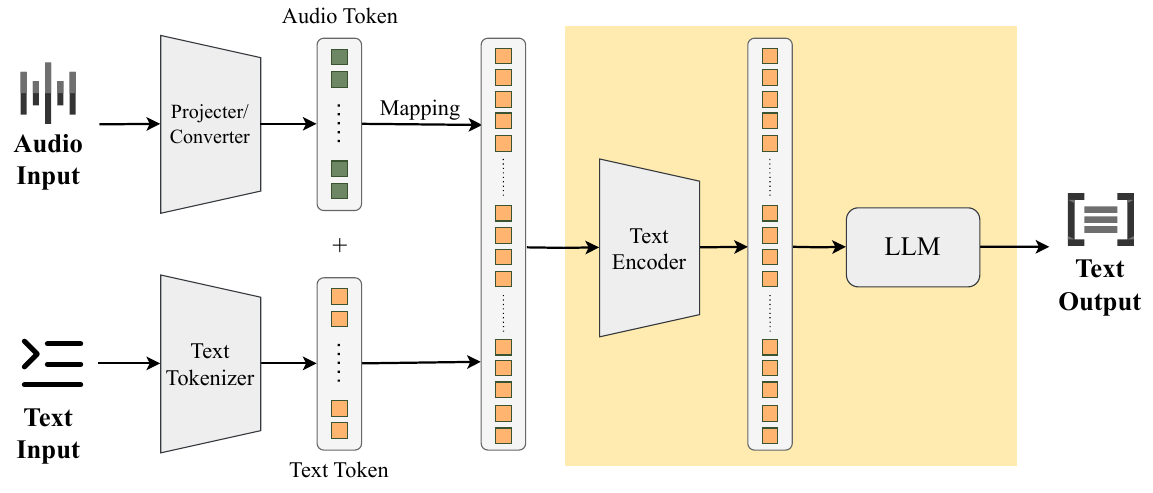}
        \caption{Modality Fusion of Discrete Embedding: Token Mapping}
        \label{fig:token_mapping}
    \end{subfigure}
    
    \vspace{1em} % Adds vertical space between the images
    
    \begin{subfigure}[t]{\textwidth}
        \centering
        \includegraphics[width=0.85\textwidth]{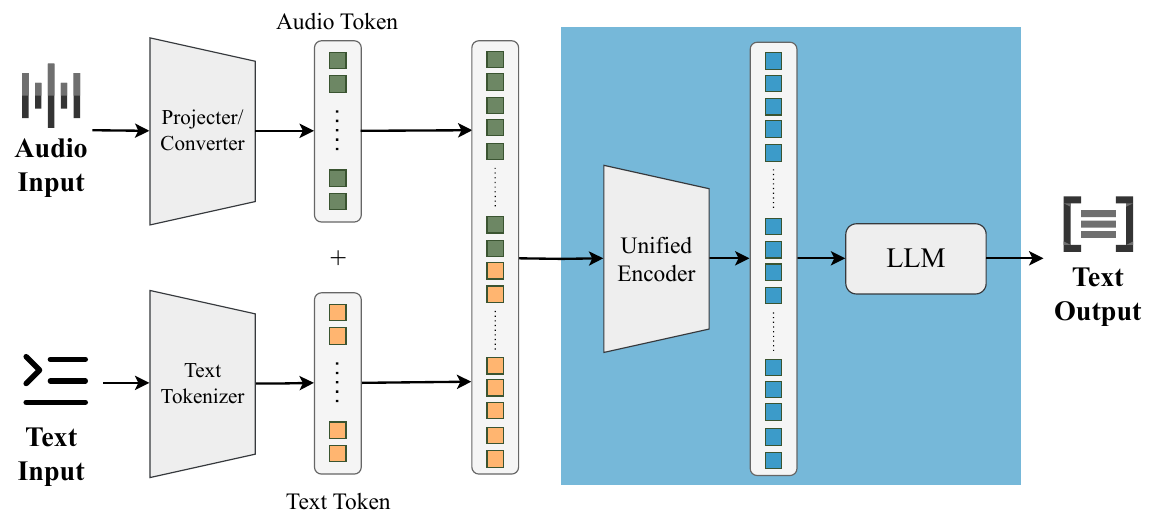}
        \caption{Modality Fusion of Discrete Embedding: Token Space Expansion}
        \label{fig:space_combine}
    \end{subfigure}

    \begin{subfigure}[t]{\textwidth}
        \centering
        \includegraphics[width=0.9\textwidth]{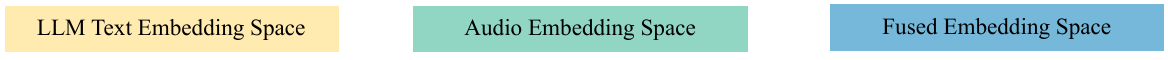}
        \label{fig:space_color}
    \end{subfigure}

    \caption{Illustration of three approaches to audio-text information fusion in Speech LLMs. The unified encoder in \ref{fig:space_combine} is transformed from the text encoder by adding audio tokens to the existing text token vocabulary.}
    \label{fig:overall}
\end{figure*}

\subsection{Modality Information Fusion}
\label{sec:fusion}
 After completing the audio feature extraction stage, the most critical challenge in Speech LLMs lies in the alignment between the audio modality and the text modality. In this section, we discuss alignment methods in both continuous and discrete fashions. As introduced in the previous section, the output formats of these two methods differ, thereby requiring customized approaches to modality fusion in each case.

 \subsubsection{Modality Fusion of Continuous Audio Embedding}
Before addressing how to effectively combine the two types of information, one important consideration is determining which specific part of the audio modality should be used. Researchers currently tend to use the output from the final layer of the encoder as the primary source of audio modality information. However, various alternative methods also exist. For example, some approaches utilize intermediate layer outputs to capture more granular features~\cite{yang2021superb}, while others apply attention mechanisms to emphasize relevant parts of the audio signal~\cite{chen2022wavlm}.

Once the audio embedding is obtained, it is merged into the text embedding space using two main methods:

\begin{enumerate}
    \item \textbf{Pre-Decoder Alignment}: This category refers to methods that align audio features with text embeddings before feeding them into the LLM decoder. A common approach is to project the audio feature information into the LLM’s text feature space through a connector~\cite{ma2024,tang2023salmonn}. Specifically, the tensor containing audio features typically has its own distinct feature dimension. The audio feature embedding $\mathbf{Z}_a$ is projected into a vector with the same dimensionality as the text embedding using a projection function $w(\cdot)$: $\mathbf{H}_a = w(\mathbf{Z}_a)$, where $\mathbf{H}_a$ denotes the projected audio embedding in the text embedding space. These audio embeddings are then concatenated with the input text embedding $\mathbf{H}_t$ to form a new embedding sequence that integrates both speech and text information: $\mathbf{H} = [\mathbf{H}_a, \mathbf{H}_t]$. This combined sequence is then fed into the LLM. Some researchers have also implicitly incorporated the projection step within the original encoder, achieving modality projection by adjusting the encoder's parameters during training~\cite{chu2023qwen}. Most models use a multi-layer perceptron (MLP) as the projection module, while some adopt more complex techniques such as Q-Former, which introduces trainable query tokens that attend to audio features and output fixed-length embeddings aligned with the LLM input space~\cite{tang2023salmonn}.
    
    \item \textbf{In-Decoder Alignment (Cross Attention)}: In addition to merging modalities before inputting them into the language model, another method establishes a bidirectional interaction between audio and text features via cross-attention~\cite{kong2024audio}. In this approach, the model maintains separate audio and text streams, allowing each modality to attend to the other through cross-attention layers, thereby enabling deeper multimodal fusion within the transformer architecture.

\end{enumerate}

\subsubsection{Modality Fusion of Discrete Audio Tokens}
For discrete audio tokens, they are typically treated in the same way as text tokens. The core challenge lies in how to process audio tokens using only a text-based vocabulary. There are essentially two ways to address this issue:

\begin{enumerate}
    \item \textbf{Token Mapping}: The most intuitive approach is to map audio tokens to text tokens that the LLM can process. Tsunoo et al.~\cite{Tsunoo2024} propose using CTC predictions to efficiently generate prompts—referred to as CTC prompts. After audio features are converted into text tokens using CTC, these tokens are merged with the tokenized input text to create a unified token sequence that represents both modalities. This sequence is then fed into the LLM for processing. This approach not only preserves the semantic content of the audio features but also ensures consistency in the LLM’s processing pipeline.

    \item \textbf{Token Space Expansion}: \label{sec:structure B} While projecting the speech modality into the text modality is straightforward, it does not achieve truly lossless modality fusion. Information loss and conflicts may occur during the modality conversion process (to be elaborated later). Therefore, researchers have proposed an alternative approach that modifies the input space of the LLM to natively incorporate the audio modality~\cite{rubenstein2023audiopalm,zhang2023speechgpt,zhan2024anygpt}. Specifically, this method augments the token space by adding audio tokens to the existing text token vocabulary, forming a unified token space. These new audio tokens are synthesized from the audio features extracted in the previous stage, thereby preserving a greater portion of the original audio information (illustrated in~\Cref{fig:space_combine}).
\end{enumerate}

\subsubsection{Modality Alignment Strategies Discussion}
\label{sec:modality_alignment_discussion}

Building on the preceding discussion, current Speech LLMs fuse modalities (i) under continuous audio representations via pre-decoder alignment or in-decoder cross-attention, and (ii) under discrete representations via token mapping or token-space expansion. 

In practice, systems vary along several choices. Beyond fusion, a fundamental design choice is which audio representation to align. Using final-layer encodings maximizes semantic abstraction, while intermediate-layer features expose finer prosodic or phonetic cues. Attention-based pooling, learned query resamplers (e.g., Q-Former/Perceiver-style), strided subsampling, and CTC-based temporal compression are all used to control the rate of audio tokens fed to the LLM, balancing information retention and compute. It is worth noting that several recent studies have explored novel alignment paradigms. For example, LegoSLM~\cite{ma2025legoslmconnectingllmspeech} leverages LLM word embeddings to construct weighted representations of speech signals, while AlignFormer~\cite{fan2025alignformermodalitymatchingachieve} effectively simplifies speech feature sequences by operating directly on the CTC posterior distributions. In a similar spirit, TASU~\cite{peng2025tasutextonlyalignmentspeech} further exploits CTC posteriors to more effectively extract semantic information from audio and to balance the information rate between the speech and text modalities.

\subsection{LLM Inference}
\label{sec:LLM_inference}
The LLM inference stage consists mainly of a text decoder, which is typically an autoregressive transformer decoder that generates text tokens conditioned on input audio and/or text. Architecturally, it often mirrors standard language models like GPT, but it is specially designed or adapted to handle multimodal inputs—such as projected audio embeddings or cross-attended audio features—alongside text tokens. To accommodate modality fusion, the decoder may incorporate special tokens (e.g., \textless aud\textgreater, \textless text\textgreater) or use prompt formatting to distinguish and contextualize different modalities during generation. Despite these adaptations, the decoder follows the same autoregressive objective: predicting the next text token given previous tokens and any additional input context. In models that generate both speech and text as output, the decoder may also be extended to support the generation of discrete audio tokens. One common approach, as seen in VALL-E, treats speech generation as a conditional codec language modeling task, where the model generates audio tokens conditioned on phoneme sequences and an acoustic prompt ~\cite{wang2023neural, chen2024vall}. Another approach used in SpeechGPT expands the LLM’s vocabulary to include both text and speech tokens, enabling the model to generate audio token sequences directly alongside or following text, depending on the instruction~\cite{zhang2023speechgpt}. In both cases, the generated audio tokens are passed through a neural codec decoder (e.g., EnCodec\cite{defossez2022high} or HiFi-GAN\cite{kong2020hifi}) to reconstruct the final waveform.

\section{Training Strategies}
\label{sec:train}
\begin{table*}[t]
\small
\renewcommand{\arraystretch}{1.3}
\centering
\caption{The following table shows the training strategy of Speech LLM. We show the training parameters at different stages, corresponding to the training data task type (as indicated inside the parentheses) and the type of capabilities the final model has. The -- symbol indicates that the paper does not mention it in detail, while \ding{56} symbol indicates that the model does not have this stage.}
\label{tab:train_strategy}
\begin{tabular}{lp{3.8cm}p{3cm}ll}

\toprule
\textbf{Model} & \textbf{Modality Alignment} & \textbf{Multitask Training} & \textbf{Preference Alignment} & \textbf{Ability} \\
\midrule
SLAM-ASR\cite{ma2024embarrassingly} & Adapter (ASR) & \ding{56} & \ding{56} & \multirow{2.8}{*}{ASR} \\
\cline{1-4}
FireRedASR\cite{xu2025fireredasropensourceindustrialgrademandarin} & Encoder + Adapter + LoRA (ASR) & \ding{56} & \ding{56} &  \\
\hline
Mala-ASR\cite{yang2024mala} & Adapter / Adapter + LoRA (ASR/Contextual ASR) & \ding{56} & \ding{56} & \multirow{2.8}{*}{Contextual ASR}  \\
\cline{1-4}
Seed-ASR\cite{bai2024seed} & Encoder + Adapter (ASR) & (Contextual ASR) & RL & \\
\hline
SLAM\cite{bapna2022slam} & Encoder + Adapter + LoRA (ASR + AST) & \ding{56} & \ding{56} & \multirow{2}{*}{ASR + AST} \\
\hline
OSUM\cite{geng2025osumadvancingopenspeech} & Encoder + Adapter + LoRA (ASR-Multitask) & \ding{56} & \ding{56} & ASR-Multitask \\
\hline
Desta\cite{lu24c_interspeech} & Adapter + LoRA (AAC) & \ding{56} & \ding{56} & \multirow{7}{*}{Multitask Speech Ability}  \\
\cline{1-4}
Qwen-audio\cite{chu2023qwen} & -- & Encoder + Adapter (Multitask) & \ding{56} &  \\
\cline{1-4}
SALMONN\cite{tang2023salmonn} & Encoder + Adapter (ASR + AAC) & Encoder + Adapter (Multitask) & \ding{56} &  \\
\cline{1-4}
Qwen2-audio\cite{chu2024qwen2} & -- (ASR + AAC) & (Multitask) & DPO &  \\
\bottomrule
\end{tabular}
\captionsetup{font=footnotesize}
\caption*{\textsuperscript{1} ASR and AAC tasks are automatic speech recognition and audio captioning tasks.\\
\textsuperscript{2} ASR-Multitask task means that the model first completes the speech recognition task before performing other tasks. \\
\textsuperscript{3} The specific tasks designed by Multitask in different papers vary greatly. Please refer to the original papers.\\
\textsuperscript{4} Context ASR means that Speech LLM supports speech recognition with the assistance of hot words or context information.\\
Although multiple parts can be trained at a certain stage, this does not mean that the model always unfreezes multiple parts at the same time. For example, the unfreezing order of OSUM is Adapter, Encoder, Lora. Please refer to the original paper for details.\\
}
\end{table*}

To support the development of understanding-oriented capabilities, Speech LLMs require carefully designed training strategies. Since most Speech LLMs designed for speech understanding rely on continuous sequence modeling, this section begins by describing training strategies specifically for models that utilize continuous embeddings. Training these models typically proceeds through three major stages: \textit{Modality Alignment}, \textit{Multitask Training}, and \textit{Preference Alignment}, as summarized in Table \ref{tab:train_strategy}. Each stage plays a distinct role in gradually aligning modalities, expanding task capabilities, and fine-tuning the model to follow human intent. However, not all models necessarily undergo all three stages; the actual training stages employed depend on each model's intended capabilities. The organization of this section will primarily follow these three stages of training for Speech LLMs with continuous sequence modeling. At the end of this section, we briefly discuss how models based on discrete speech tokens differ in their training approach.

\subsection{Modality Alignment}
\label{sec:alignment}
At the outset, Speech LLMs must learn a shared representation between spoken input and text. In this \emph{modality alignment} phase, models typically leverage vast speech data to bridge the audio-text gap. A common strategy is to incorporate a pre-trained speech encoder and align its acoustic representations with a text-based LLM. For example, Qwen-Audio begins with a pretrained Whisper encoder coupled to an LLM backbone via a projector~\cite{qwen2025qwen25technicalreport}. The alignment is refined by supervised tasks that force audio and text to correspond. Notably, automatic speech recognition (ASR) and audio captioning (AAC) are usually used as fundamental alignment tasks~\cite{lu2025desta2developinginstructionfollowingspeech,bai2024seed}. 

The total parameters $\theta$ of the model are composed of $\theta_{\text{LLM}}$, $\theta_{\text{encoder}}$, and $\theta_{\text{adapter}}$ . The model is trained to map an input audio sequence $a$ to a textual output $y$ (transcript or caption). The training data consist of many triples $(a, p, y)$: an audio input $a$, a textual prompt or task descriptor $p$, and the expected textual output $y$. The model is optimized to predict $y$. In this stage, the textual prompt $p$ typically remains unchanged, which makes modality alignment easier to train \cite{fathullah2024prompting,li2024transcription}. The training loss can be represented as 
% , effectively minimizing a cross-entropy loss 
\begin{equation}
\label{equ:align}
\mathcal{L}_{\text{align}}(\theta) = -\log P(y \mid \theta_{\text{LLM}}(\theta_{\text{adapter}}(\theta_{\text{encoder}}(a) ), p).
\end{equation}
Depending on the training objectives, many models focus on a single speech-to-text task (e.g., speech recognition) rather than training a multitask speech model capable of fully understanding speech. Therefore, many models only have one stage of modality alignment. The most common approach is to train only the adapter \cite{yu2024connecting,li2024transcription}. However, after the initial alignment, continuing to add LoRA parameters for further training has been shown to maintain better modality alignment and achieve superior alignment performance \cite{fathullah2024prompting}. Another training approach involves simultaneously training the encoder, adapter, and LoRA parameters, as seen in SLAM and FireRedASR \cite{chen2024salm,xu2025fireredasropensourceindustrialgrademandarin}. Interestingly, both models employ Conformer as the encoder, and the FireRedASR model has achieved SOTA performance on current recognition leaderboards—perhaps demonstrating that unfreezing the encoder and training with more parameters can yield better alignment results.
There is currently no unified consensus on the order of training parameters. A common practice is to first train the adapter, then train the encoder and adapter, and finally add LoRA parameters for joint training \cite{geng2025osumadvancingopenspeech}. However, it is also advisable to include stages where some parameters are frozen during training \cite{xu2025leveraging}. For multitask models like Qwen-audio \cite{chu2023qwen} and Seed-ASR \cite{bai2024seed}, LoRA parameters of LLMs are typically not used as training parameters for modality alignment—partly to preserve the original powerful contextual and reasoning abilities of the LLMs, expecting them to play a role in subsequent multitask training. Not all models follow this pattern, however: SALMONN \cite{tang2023salmonn} trains LoRA parameters during the alignment phase.
 % This is evidenced by improved transcription and audio comprehension early in training—OSUM, for instance, reports that concurrently training an ASR task alongside any new audio task (an “ASR+X” strategy) accelerates alignment, as the model always learns to extract and decode the spoken words before handling additional subtasks. Overall, this phase imbues the model with a basic “speech understanding” ability, aligning phonetic and lexical content of audio with text tokens, which is crucial before tackling more complex or diverse tasks.

\subsection{Multitask Training}
\label{sec:multitask}
After modality alignment, speech features are mapped into the text space of the LLM. However, even when an LLM can accurately transcribe speech, it often loses the ability to reason about or follow instructions related to the transcribed content. Therefore, the LLM must undergo a multitask training phase to become a multitask speech model.

Taking models focused solely on ASR tasks as an example, during this phase many models add LoRA parameters to leverage the LLM's contextual capabilities, including hotword usage and historical transcripts, to improve recognition results \cite{lakomkin2024end,bai2024seed}. However, these are not standard multitask training approaches but rather extensions of ASR tasks.

According to the requirements of different models for speech understanding tasks, there is no unified standard for the division of multitask content. SALMONN~\cite{tang2023salmonn} classifies speech tasks into three difficulty levels: the first level is simple tasks, the second level is speech-based NLP tasks such as keyword extraction and translation, and the third level is speech-based reasoning tasks such as speech-based question answering and storytelling. OSUM~\cite{geng2025osumadvancingopenspeech} mainly focuses on tasks such as audio emotion classification and audio style classification. It is worth noting that although this paper mainly focuses on the speech understanding tasks defined by us, some models have extended speech capabilities to sounds and music, aiming to train a general Speech LLM. For example, Qwen-Audio \cite{chu2023qwen} involves multitask training related to speech, sound, and music simultaneously.

The mainstream training approach resembles modality alignment: natural language instructions are concatenated with the aligned speech features 
 and target text, and the model is directly trained for multitask via cross-entropy loss. Generally, LoRA parameters are added to the training at this stage to fine-tune the LLM, aiming to enable the LLM to acquire the multitask speech capability.

Notably, Qwen-Audio \cite{chu2023qwen} employs a Whisper-like multitask training framework that uses task prefixes and does not use LoRA for fine-tuning. In contrast, OSUM \cite{geng2025osumadvancingopenspeech} uses an \textbf{ASR+X} approach, where \textbf{X} represents other tasks. This means OSUM transcribes speech into text before performing any other audio tasks. 

\subsection{Preference Alignment}
\label{sec:preference_alignment}
The final development phase focuses on aligning the model’s behavior with user instructions and preferences, ensuring that it follows natural language commands reliably and produces responses that are helpful and context-appropriate. 

After multitask alignment, some advanced Speech LLMs apply \emph{preference alignment} using reinforcement learning ~\cite{qwen2025qwen25technicalreport}. This step optimizes the model’s responses according to human feedback or a learned reward function, refining qualities such as factual accuracy, relevance, or user satisfaction~\cite{chen2024enhancing,tian2024preference}, for example, Seed-ASR\cite{bai2024seed}. A prominent method is reinforcement learning from human feedback (RLHF), operationalized via Proximal Policy Optimization (PPO)~\cite{schulman2017proximal}. In RLHF, the model is treated as a policy $\pi_\theta(y|X)$ that generates a response $y$ to an input $X$, and a reward model $R(X,y)$ (trained from human preference rankings) scores the response. The goal is to maximize expected reward $E_{X}[E_{y\sim \pi_\theta}[R(X,y)]]$ while keeping the language model’s output distribution close to the pre-trained model to avoid degrading fluency. PPO achieves this by updating $\theta$ in the direction of the policy gradient $\nabla_\theta J \approx E_{X,y}[R(X,y)\,\nabla_\theta \log \pi_\theta(y|X)]$ with constraints on the KL-divergence from the original policy. In practice, the PPO update uses a clipped surrogate objective to ensure stable improvements. An alternative, lightweight approach is \emph{Direct Preference Optimization} (DPO)~\cite{rafailov2024directpreferenceoptimizationlanguage}, which was employed in Qwen2-Audio’s final training. Unlike methods that require training a separate reward model, DPO directly leverages preference pairs \((y^+, y^-)\) for a given prompt \(x\). It optimizes the policy to assign higher probability to the preferred response \(y^+\) over the less favored \(y^-\). The DPO loss can be formulated as:

\begin{equation}
\label{equ:dpo}
\resizebox{0.91\hsize}{!}{
$\mathcal{L}_{\text{DPO}}(\theta) = -\mathbb{E}_{(x,y^+,y^-) \sim \mathcal{D}} \left[ \log \sigma\!\left(\frac{1}{\beta}\big(\log \pi_\theta(y^+|x) - \log \pi_\theta(y^-|x)\big)\right) \right]\!,$
}
\end{equation}

where \(\sigma\) is the sigmoid function and \(\beta\) is a temperature hyperparameter. By minimizing this loss, the model is encouraged to prefer \(y^+\) over \(y^-\), effectively aligning its outputs with human preferences without the complexity of a reward model training loop. By the end of this reinforcement or preference alignment phase, the Speech LLM is highly tuned to follow instructions in a human-preferred way: it will listen to a user’s voice request and respond with appropriate content. Crucially, this stage often improves the model’s \textit{factuality} and safety by penalizing incorrect or undesirable continuations. Qwen2-Audio’s authors note that after DPO fine-tuning, the model more faithfully adheres to user intent and makes fewer mistakes in complex audio-centric instructions. The strength of the instruction alignment phase lies in making the model not just capable, but also reliably usable: it transforms the multitask-trained foundation into a polite conversational agent that can handle open-ended voice interactions. However, it is worth noting that this phase has received limited attention in current Speech LLM research focused on speech understanding. Most existing works primarily emphasize modality alignment and task-specific performance, while instruction-following behavior, dialogue consistency, and safety alignment remain underexplored. As Speech LLMs continue to evolve toward general-purpose audio agents, further investigation into instruction and preference alignment is expected to play a crucial role in improving usability, controllability, and user interaction quality.

 All of these stages combined – modality alignment, broad multitask learning, and final alignment with preferences – produce state-of-the-art Speech LLMs that can understand speech in rich context and interact with users naturally, bridging the gap between spoken language and the powerful reasoning of text-based LLMs.

 \subsection{Training of Speech LLMs with Discrete Sequence Modeling}
 \label{sec:discrete_training}
In contrast to the three-stage training pipeline of continuous embedding models, Speech LLMs based on discrete speech tokens adopt a simpler and more unified training strategy. For these models (e.g., AudioPaLM~\cite{rubenstein2023audiopalm}, SpeechGPT~\cite{zhang2023speechgpt}, VALL-E~\cite{wang2023neural}), after generating the discrete tokens for audio representations, these token sequences are treated analogously to text tokens, allowing a pretrained or newly initialized language model to be trained directly using a standard autoregressive objective.

Training typically proceeds by fine-tuning on paired speech-text data in a unified token space, without the need for a separate modality alignment stage, since discrete-token models do not introduce structural modifications such as adapters, which are common in continuous embedding models. This simplifies the pipeline and enables models to benefit directly from pretrained LLMs. In practice, alignment between modalities is learned implicitly through multitask fine-tuning, where the model learns to interleave and reason over both speech and text tokens. For example, SpeechGPT~\cite{zhang2023speechgpt} introduces a “Paired Speech-Text Pretraining” phase to warm up the model on interleaved sequences, followed by instruction fine-tuning using synthetic multimodal dialogues. Unlike continuous-embedding Speech LLMs, mainstream discrete-token Speech LLMs typically skip explicit preference alignment (e.g., RLHF or DPO). Their training focuses on supervised multitask fine-tuning over speech–text data without a separate reward modeling stage. While emerging research such as SpeechAlign~\cite{zhang2024speechalign} has begun to explore preference optimization for codec-based language models, these methods remain experimental and are not yet part of the standard speech-LLM training toolbox.

This training paradigm offers the benefit of architectural simplicity and efficient reuse of existing LLM infrastructure. However, it comes with its own challenges, such as information loss from quantization and reduced acoustic fidelity.

\section{Datasests}
\label{sec: Datasets}
\begin{table*}[t]
    \centering
    \caption{Overview of commonly used speech datasets categorized by task type from the perspective of functional dimension. It spans various domains including perception tasks and shallow cognition tasks which are typically based on speech dataset in traditional format.}
    \label{table:dataset}
    \renewcommand{\arraystretch}{1.2}
    \newcolumntype{S}{>{\small}c}
    
    \begin{tabular}{l l l l l}
        \toprule
        \textbf{Task Type} & \textbf{Dataset Name} & \textbf{Covered Languages} & \textbf{Total Duration/Dataset Size} & \textbf{Year} \\
        \midrule
        \multicolumn{5}{c}{\textbf{Perception Tasks}} \\
        \midrule
        \multirow{12}{*}{Automatic Speech Recognition} 
            & LibriSpeech \cite{panayotov2015librispeech} & English & $\sim$1000 hours & 2015 \\
        \cmidrule(lr){2-5}
            & Aishell1 \cite{bu2017aishell} & Mandarin Chinese & 170 hours & 2016 \\
        \cmidrule(lr){2-5}
            & Common Voice \cite{ardila2019common} & 100+ Langs & 33000+ hours & 2017 \\
        \cmidrule(lr){2-5}
            & Aishell2 \cite{du2018aishell} & Mandarin Chinese & 1000 hours & 2018 \\
        \cmidrule(lr){2-5}
            & TED-LIUM 3 \cite{hernandez2018ted} & English & 452 hours & 2018 \\
        \cmidrule(lr){2-5}
            & MLS \cite{pratap2020mls} & 8 Langs & $\sim$50k hours & 2020\\
        \cmidrule(lr){2-5}
            & Aishell3 \cite{shi2020aishell} & Mandarin Chinese & 85 hours & 2020\\
        \cmidrule(lr){2-5}    
            & VoxPopuli \cite{wang2021voxpopuli} & English + 14 EU Langs & $\sim$1000 hours/lang & 2021 \\
        \cmidrule(lr){2-5}
            & GigaSpeech \cite{chen2021gigaspeech} & English & $\sim$10000 hours & 2021 \\
        % \cmidrule(lr){2-5}
        %     & KeSpeech\cite{tang2021kespeech} & Mandarin Chinese & 1500+ hours& 2021 \\
        \midrule
        \multirow{2.4}{*}{Speaker Diarization}
            & AMI Meeting Corpus \cite{kraaij2005ami} & English & 100 hours & 2005 \\
        \cmidrule(lr){2-5}
            & VoxConverse \cite{chung2020spot} & Multilingual & 50+ hours & 2020 \\
        \midrule
        \multirow{2.4}{*}{Speaker Identification/verification}
            & VoxCeleb2 \cite{chung2018voxceleb2} & English & 2,442 hours & 2018 \\  
        \cmidrule(lr){2-5}
            & CN-Celeb \cite{fan2020cn} & Chinese, English & $\sim$700 hours & 2020 \\
        \midrule
        \multirow{2.4}{*}{Keyword Spotting}
            & Mobvoi Hotwords \cite{hou2019region} & Mandarin Chinese & $\sim$220 hours & 2019 \\
        \cmidrule(lr){2-5}
            & Hey Snips \cite{coucke2019efficient} & English & $\sim$3 hours & 2019 \\
        \midrule
        \multirow{2.4}{*}{Voice Activity Detection}
            & AVA-Speech \cite{chaudhuri2018ava} & Multilingual & $\sim$45 hours & 2018 \\
        \cmidrule(lr){2-5}
            & VoxConverse (VAD) \cite{chung2020spot} & Multilingual & 50+ hours & 2020 \\
        \midrule
        Vocal Sound Classification & VocalSound\cite{gong2022vocalsound} & Multilingual & 21000+ & 2022\\
        \midrule
        \multicolumn{5}{c}{\textbf{Shallow Cognition Tasks}} \\
        \midrule
        \multirow{6.4}{*}{Emotion Classification}
            & IEMOCAP \cite{busso2008iemocap} & English & $\sim$12 hours & 2008 \\
        \cmidrule(lr){2-5}
            & MELD\cite{poria2018meld} & English & 13000+ utterances & 2018\\
        \cmidrule(lr){2-5}   
            & RAVDESS \cite{livingstone2018ryerson} & English & $\sim$2.4 hours & 2018 \\
        \cmidrule(lr){2-5}
            & MSP-Podcast \cite{martinez2020msp} & English & 100 hours & 2020 \\
        \cmidrule(lr){2-5}
            & ESD \cite{zhou2022emotional} & English, Chinese & $\sim$29 hours & 2021 \\
        \midrule
        \multirow{3.8}{*}{Speech-to-Text Translation}
            & MuST-C \cite{di2019must} & En $\rightarrow$ 8 & 385–500 hours/lang & 2019 \\
        \cmidrule(lr){2-5}
            & CovoST2 \cite{wang2021covost} & 21 $\rightarrow$ En; En $\rightarrow$ 15 & 2,880 hours & 2020 \\
        \cmidrule(lr){2-5}
            & FLEURS \cite{conneau2023fleurs} & 102 Langs & $\sim$12 hours/lang & 2022 \\
        % \midrule
        % \multirow{3}{*}{Speech Summarization}
        %     & AMI Meeting Corpus \cite{kraaij2005ami} & English & 100 hours & 2005 \\
        % \cmidrule(lr){2-5}
        %     & How2 (Multimodal) \cite{sanabria18how2} & English & 2,000 hours & 2018 \\
        % \cmidrule(lr){2-5}
        %     & Spotify Podcasts \cite{clifton2020spotify} & English & 47,000 hours & 2020 \\
        % \midrule
        % Task-Oriented SLU & SLURP \cite{bastianelli2020slurp} & English & 58 hours & 2020 \\
        \bottomrule
    \end{tabular}
\end{table*}

In parallel with model advancements, the availability and design of datasets have also evolved to support increasingly complex speech understanding tasks. In this section, we categorize and analyze these datasets through the lens of their respective understanding objectives, as defined in \ref{taxonomy:functional}. We categorize speech understanding tasks according to their objectives, from simpler and more immediate tasks, such as perception tasks, to more complex and deeper objectives that require advanced reasoning, including shallow and deep cognition tasks. This progression parallels the evolution of speech models, which initially focused on handling fundamental tasks but have since advanced to address more intricate cognitive functions. Here, we provide a detailed overview of the various categories of speech datasets and examine how they have evolved, particularly in response to the development of LLMs and TTS.

\subsection{Datasets for Perception and Shallow Cognition Tasks}
\label{sec:traditional_dataset}

Perception and shallow cognition tasks focus on extracting information from audio signals and performing intuitive understanding, without requiring deep semantic reasoning. Tasks such as Automatic Speech Recognition (ASR), Speaker Diarization (SD), and Keyword Spotting (KWS) are examples of perception tasks, while emotion recognition and speech translation are considered shallow cognition tasks. These types of tasks correspond to traditional single-task datasets that predate the era of LLMs. A significant portion of the datasets used to train Speech LLMs consists of audio-text pairs, which are readily available from many traditional single-task datasets. In addition to these audio-text pairs, tasks like speaker diarization and keyword spotting often include additional metadata, such as timestamps and speaker labels, to enhance task performance. Rather than being limited to single-task datasets, multi-task datasets with rich metadata are also widely used, enabling training across multiple tasks. The AMI Meeting Corpus is a representative example of such datasets, which can be used for a variety of tasks as shown in \cite{carletta2005ami, Yu2022M2MeT,Yu2022Summary, fu2021aishell}. These datasets have become foundational in speech understanding due to their early development and wide availability. In addition to being used for pre-training Speech LLMs, these datasets can also serve as direct resources for downstream tasks, facilitating the fine-tuning of Speech LLMs.

Table~\ref{table:dataset} categorizes datasets associated with perception and shallow cognition tasks, including ASR, speaker analysis, KWS, emotion classification, and speech translation. This categorization highlights the diversity of task types, language coverage, and dataset scale, reflecting the fundamental role of these resources in the broader field of speech understanding.

\begin{table*}[t]
\centering
\caption{Overview of publicly available speech-based understanding datasets. The table categorizes datasets into three groups: SLU, Question Answering about Spoken Content, and Spoken Question Answering.}
\label{table:speech_qa}
\renewcommand{\arraystretch}{1.2}
\begin{tabular}{p{2.7cm}p{3.8cm}p{2.5cm}lp{2.7cm}l}
\toprule
\textbf{Dataset Name} & \textbf{Input Modality} & \textbf{Output} & \textbf{Languages} & \textbf{Size / Duration} & \textbf{Year} \\
\midrule
\multicolumn{6}{c}{\textbf{Traditional SLU Tasks}} \\
\midrule

% \hline
ATIS3\cite{hemphill1990atis, dahl1994expanding} & Spoken utterance + transcript & Intent + slots & English & 11k utterances & 1990 \\
\hline
SNIPS (spoken)\cite{coucke2018snips} & Spoken command recordings & Intent + slots & English & 5.9k utterances & 2018 \\
\hline
CATSLU\cite{zhu2019catslu} & Spoken utterance & Intent classification & Chinese & 16k utterances & 2019 \\
\hline
SLURP \cite{bastianelli2020slurp} & Spoken utterance & Intent + slots& English & 72k utterances & 2020 \\
\hline
Speech-MASSIVE\cite{lee2024speech} & Spoken utterance & Intent + slots & 12 languages & 83k utterances & 2024 \\
% \hline
% CSLU: Portland\footnotemark[\value{footnote}] & Telephone speech recordings & Transcript & English & 7,571 utterances from 515 speakers & – \\
\midrule
\multicolumn{6}{c}{\textbf{SQA: Question Answering about Spoken Content}} \\
\midrule
TOEFL-QA\cite{tseng2016towards} & Spoken lecture + text question (MCQ) & Text (choice) & English & 963 QA pairs & 2016 \\
\hline
Spoken SQuAD\cite{li2018spoken} & Spoken passage + text question & Text span & English & 42k QA pairs & 2018 \\
\hline
ODSQA\cite{lee2018odsqa} & Spoken document + spoken/text question & Text span & Chinese & 7k QA pairs & 2018 \\
\hline
Spoken-CoQA\cite{you2022end} & Spoken passage + spoken questions & Text (free-form) & English & 40k QA turns & 2022 \\
\hline
NMSQA\cite{lin2022dual} & Spoken passage + spoken question & Text span & English & 92k QA pairs, 335h audio & 2022 \\
\hline
CORAAL-QA\cite{shankar2024coraal} & Spontaneous long-form audio + text query & Text span & English (AAVE) & 143 interviews, hundreds of QA pairs& 2024 \\
\hline
AIR-Bench\cite{yangAIRBenchBenchmarkingLarge2024} &  Speech/Music/Sound/Mixed + text query& Text (choice/free-form) & English  & 18.6k VMCQ + 2k QA pairs& 2024 \\
\hline
MMAU\cite{sakshi2024mmau} & Speech/Music/Sound + text query & Text (choice) & English& 10k VMCQ & 2025 \\
\hline
MMAR\cite{ma2025mmar} & Speech/Music/Sound/Mixed + text query & Text (choice) & English  & 1k VMCQ  & 2025 \\
\midrule
\multicolumn{6}{c}{\textbf{SQA: Spoken Question Answering}} \\
\midrule
SD-QA (TyDi)\cite{faisal2021sd} & Spoken question + text context & Text span & 5 languages & 68k spoken questions & 2021 \\
\hline
SQuAD-SRC\cite{tang2023squad} & Spoken question + text context & Text span & English (6 accents) & 10k spoken questions & 2023 \\
\hline
HeySQuAD\cite{wu2023heysquad} & Spoken question + text context & Text span & English & 173k spoken questions & 2024 \\
\hline
SpokenNativQA\cite{alam2025spokennativqa} & Spoken question & Text answer & English, Arabic & 33k spoken questions & 2025 \\
\bottomrule
\end{tabular}
\end{table*}

\subsection{Datasets for Deep Cognition Tasks}
\label{sec:deep_datasets}
As Speech LLMs have advanced, traditional datasets have become increasingly insufficient for training models capable of handling more complex reasoning tasks. Traditional SLU, which involves extracting structured semantic information from spoken utterances, serves as an intermediary step toward deeper cognitive tasks. Such SLU tasks typically require models to perform tasks with structured output, such as identifying intents and slot values from audio input, representing a constrained form of deep cognitive reasoning. A standard example of traditional SLU is Dialogue State Tracking (DST), as exemplified by datasets such as \textit{DialogZoo}~\cite{chen2022dfm}, where models track dialogue states to facilitate structured understanding of dialogues.

Building upon these capabilities, researchers have introduced more comprehensive datasets explicitly designed for deeper cognitive reasoning tasks, such as speech question answering (SQA), which generally demands more advanced reasoning abilities compared with traditional SLU tasks. Inspired by text-based question answering datasets from Natural Language Processing (NLP), SQA datasets typically involve providing context information and questions in speech format, requiring models to extract relevant details from the audio and subsequently generate textual answers. Some of these SQA datasets present the question in the format of multiple choice, for which we denote it as Voice-based Multiple-choice Question (VMCQ). It is exemplified by MMAR~\cite{ma2025mmar} and MMAU~\cite{ma2025mmar}, where models are demanded with more complex reasoning ability based solely on the input audio content.

These datasets necessitate that models not only extract pertinent information from speech inputs but also reason effectively using the extracted details, aligning closely with the cognitive capabilities required for deep reasoning tasks.

Table~\ref{table:speech_qa} presents a unified overview of speech-based understanding datasets, broadly divided into two parts: SLU and Speech Question Answering (SQA).  Within SQA, we further categorize datasets into two types: those that require answering questions about spoken content (e.g., lectures or conversations), and those where the question itself is spoken. This perspective highlights the growing trend toward general-purpose spoken language understanding and supports the development of unified models capable of handling speech input with textual reasoning.

\subsection{Recent Developments in Datasets Leveraging LLM and TTS Technologies}
\label{sec:recent_datasets}
Recent advances in LLMs and Text-to-Speech (TTS) technology have led to significant improvements in dataset creation and annotation for multimodal tasks. Researchers are increasingly turning to LLMs for their potential in automating and enhancing dataset annotation processes, particularly for tasks involving speech and text. In this part, we will explore how LLMs and TTS technologies are being utilized to create more scalable and diverse datasets, as well as the challenges and opportunities associated with these approaches.

\subsubsection{Datasets Annotated with LLM Participation}

The rapid advancement of LLMs, particularly in text-based reasoning capabilities approaching human performance, has inspired researchers to explore their potential in dataset annotation. This approach typically involves representing audio information in text form and pairing it with task-specific prompts to guide LLM in generating inferences. This process, in the \textit{SQA format (Speech [in text] + Question + Answer)}, allows for scalable dataset creation by reducing human labor and lowering costs. Furthermore, it provides a novel way to train LLM, facilitating general question-answering capabilities about speech that cannot be fully realized using traditional speech datasets alone. During training, the text-based question and its corresponding speech context are provided as input, while the model is trained to generate answers in text form.
% as illustrated in \ref{fig:dataset_LLM}

However, despite its efficiency, this method is limited by the reasoning capabilities of the LLM, which may lead to suboptimal annotations in many cases. The text prompts and answers generated by LLM often lack the diversity seen in human-created instructions. Consequently, such datasets may still fall short in supporting comprehensive speech question-answering tasks.

\subsubsection{Datasets Generated Using TTS Technology}

Given the scarcity of audio datasets specifically designed for Speech LLMs and their limited variety, researchers have proposed an innovative approach to integrate audio and text modalities more deeply to enable the model's capability of handling flexible mixed-modal input. Building on the SQA framework, this approach converts part of the textual components (speech/questions) into speech using Text-to-Speech (TTS) technology, thereby creating Audio-Text pairs for training Speech LLMs. 
% The structure is illustrated in \ref{fig:dataset_TTS}.

Peng et al. introduced VoiceTextBlender, marking the first significant effort to enhance mixed-modal supervised fine-tuning (SFT) data using TTS\cite{peng2024voicetextblender}. This method enables both prompts and questions to be represented in the audio modality, offering greater potential for multi-modal task training. This approach opens new possibilities for creating datasets that support deeper cross-modal integration while addressing the limitations of real-recorded audio datasets.

% \begin{figure*}[htbp] % 使用 figure 环境来放置图片
%     \centering
%     \begin{subfigure}[t]{0.45\textwidth}
%         \centering
%          \includegraphics[width=\textwidth]{images/dataset_LLM.pdf}
%         \caption{Dataset construction pipeline  leveraging LLM}
%         \label{fig:dataset_LLM}
%     \end{subfigure}
%     \hfill
%     \begin{subfigure}[t]{0.45\textwidth}
%         \centering
%          \includegraphics[width=\textwidth]{images/dataset_TTS.pdf}
%         \caption{Dataset construction pipeline based on TTS}
%         \label{fig:dataset_TTS}
%     \end{subfigure}
    
%     \caption{Two ways of generating SFT data for Speech LLMs training}
%     \label{fig:dataset} % 可选：用于引用此图片
% \end{figure*}

\section{Performance in Speech Tasks}
\label{sec: Performance}

\subsection{A Taxonomy of Evaluation Strategies for Speech Understanding Tasks}
\label{sec:taxonomy_evaluation}

To facilitate consistent and principled evaluation of Speech LLMs, we present a structured overview of existing evaluation strategies aligned with the taxonomy of speech understanding tasks. Rather than proposing a novel evaluation paradigm, this framework synthesizes widely adopted metrics and assessment techniques across structured, weakly structured, and unstructured tasks, as illustrated in Section \ref{taxonomy:output}. Specifically, we categorize evaluation approaches into two primary regimes—accurate and inaccurate—corresponding to the degree of structure in task outputs in and the availability of definitive groundtruths.

\subsubsection{Accurate Evaluation}

Deterministic evaluation pertains to structured speech understanding tasks where outputs are discrete, unambiguous, and benchmarked against predefined references or label sets. This category supports reproducible testing and facilitates rigorous model comparison. Representative tasks and their standard evaluation criteria include:

\begin{itemize}
\item
\textbf{Automatic Speech Recognition (ASR)}: Word Error Rate (WER), Character Error Rate (CER), Match Error Rate (MER)
\item
\textbf{Speaker Identification and Verification}: Accuracy, Equal Error Rate (EER), Detection Cost Function (DCF)
\item
\textbf{Speaker Diarization}: Diarization Error Rate (DER), encompassing missed speech, false alarms, and speaker confusion
\item
\textbf{Keyword Spotting (KWS)}: Precision, recall, F1-score, detection error trade-off (DET) curves
\item \textbf{Voice Activity Detection (VAD)}: Frame-level accuracy, false acceptance/rejection rates
\item \textbf{Emotion and Intent Classification}: Accuracy, macro/micro F1-score, confusion matrices
\item \textbf{Task-Oriented Structured Tagging}: Slot F1, intent accuracy, frame-level accuracy
\end{itemize}

These deterministic metrics are especially suitable for safety-critical, resource-constrained, or latency-sensitive scenarios where model behavior must be predictable and auditable.

\subsubsection{Inaccurate Evaluation}

Open-ended evaluation is applied to weakly structured or unstructured tasks, where groundtruths are flexible and diverse, and multiple outputs may be equally valid. This regime assesses generative quality, semantic alignment, and contextual appropriateness using a combination of similarity metrics and learned evaluation models.

Common approaches include:

\begin{itemize}
\item \textbf{Reference-based Similarity Metrics}:
\begin{itemize}
\item BLEU~\cite{papineni-etal-2002-bleu}: Measures n-gram precision, primarily used in translation.
\item ROUGE~\cite{lin2004rouge}: Emphasizes recall in summarization tasks.
\item METEOR~\cite{banerjee2005meteor}: Incorporates synonymy, stemming, and word alignment.
\item BERTScore~\cite{zhang2020bertscoreevaluatingtextgeneration}: Computes contextual embeddings similarity with pretrained models.
\end{itemize}

\item \textbf{Model-based or Human Judgments}:
\begin{itemize}
    \item LLM-based Scoring: Employs pretrained language models to assess fluency, coherence, and faithfulness~\cite{fu2023gptscore, liu-etal-2023-geval}.
    \item Human Evaluation: Includes Likert-scale ratings, pairwise preferences, or rubric-based reviews for naturalness and appropriateness.
    \item Task-Specific Rubrics: Particularly useful for qualitative assessments like speaker description, topic abstraction, or emotion narration.
\end{itemize}

\end{itemize}

These methods are indispensable in capturing subjective nuances, creativity, and interpretive reasoning that fall outside the scope of traditional classification.

It is important to note that the categorization of evaluation does not rigidly follow the task definition, but rather depends on how a task is operationalized. For example, the task of understanding emotion may be:

\begin{itemize}
\item Framed as a classification problem --- evaluated using deterministic metrics like F1-score.
\item Framed as a free-form description --- assessed via BERTScore, LLM evaluations, or human ratings.
\end{itemize}

This flexibility highlights the necessity for adaptive evaluation strategies that align with the prompt format and intended output structure, particularly as Speech LLMs adopt natural language instruction-based paradigms.

In summary, our dual-track evaluation framework accommodates structured, unstructured, and weakly-structured tasks, fostering comprehensive and principled analysis of Speech LLMs across diverse speech understanding scenarios.
To further ground this framework in practice, the following sections present a detailed analysis of representative tasks from each category—namely, ASR as a canonical structured task, ST as a prototypical weakly-structured task, and multitask capabilities as a hallmark of unstructured understanding.

% With the evolution of Speech LLMs, the integration of large language models (LLM) into speech-based systems has yielded substantial advancements across various dimensions. This section investigates the current performance of Speech LLMs in several pivotal tasks in speech understanding, evaluating how LLM enhance the capabilities of speech models compared to traditional methods. We will primarily focus on two classic speech-related tasks: \textbf{Automatic Speech Recognition(ASR)}, and \textbf{Speech Translation(ST)}, each epitomizing a crucial aspect of speech applications. Meanwhile, we will further explore the \textbf{multitasking} and \textbf{cross-tasking} capabilities exhibited by Speech LLMs across various tasks in the speech understanding.
\subsection{Performance of Speech LLMs Across Tasks}
\label{sec:performance_LLM}

Following the taxonomy of evaluation strategies for speech understanding tasks, in this section, we quantitatively analyze the current progress of Speech LLMs by evaluating their performance across four representative speech tasks. These tasks span all linguistic, paralinguistic, and non-linguistic aspects of spoken language, corresponding to the informational dimension outlined in our taxonomy of speech understanding in Section \ref{taxonomy:information}. 

\subsubsection{Analysis Setup}
Specifically, we select four tasks that together reflect the breadth of speech understanding: two tasks that primarily target linguistic information, one that focuses on paralinguistic cues, and one that emphasizes non-linguistic acoustic context.

The linguistic tasks evaluated in this section are:
\begin{itemize}
    \item Automatic Speech Recognition (ASR), where the task is to transcribe spoken language into written text. We use the LibriSpeech dataset, which includes the widely used test-clean and test-other subsets. These are among the most well-known and frequently reported datasets in the evaluation of ASR models.
    \item Speech Translation (ST), where the model translates spoken language from one language to another. For this task, we utilize the CoVoST2-En2Zh dataset, a benchmark for speech-to-text translation from English to Chinese. This dataset is recognized for its comprehensive coverage and frequent use in recent speech LLM reports.
\end{itemize}

The paralinguistic task evaluated is:
\begin{itemize}
    \item Emotion Recognition, where the model detects and classifies emotions from speech. We focus on the MELD dataset, a popular dataset for emotion recognition, which captures emotions expressed in dialogue. MELD is widely used in studies on emotion recognition and is often referenced in the context of speech LLM evaluation.
\end{itemize}

The non-linguistic task evaluated is:
\begin{itemize}
    \item Human Sound Event Classification, where the task is to identify various human vocal sounds, such as laughter, sighs, and coughs. For this task, we use the VocalSound dataset, a specialized collection for classifying non-speech human vocalizations. It is one of the most cited datasets in reports on human sound event classification.
\end{itemize}

These four datasets—LibriSpeech, CoVoST2, MELD, and VocalSound—are the most widely used in their respective tasks and are frequently referenced in the reports of the speech LLM models we have reviewed. By evaluating Speech LLMs on these datasets, we aim to gain a clearer understanding of their current capabilities and limitations across a diverse range of speech understanding tasks.

When evaluating the performance of models, we use a metric that quantifies ``how much a model can achieve compared to the State-of-the-Art (SOTA)". Formally, this metric is designed to provide a relative comparison of the model’s performance against the best-known performance (SOTA) on the task. We denote this metric as \textit{RPS} (Relative Performance to State-of-the-Art).

For metrics where larger values are better (such as accuracy, F1-score, etc.), the evaluation metric is defined as:
\begin{equation}
\text{RPS} = 
\frac{\text{Model\ Score}}{\text{SOTA\ Score}}.
\end{equation}
For metrics where smaller values are better (Word Error Rate (WER) specifically), the normalized metric is defined as:
\begin{equation}
\text{RPS} = 
\frac{\text{SOTA\ Score}}{\text{Model\ Score}}.
\end{equation}

By using these normalized values, we can uniformly compare model performance across tasks where different evaluation metrics may have opposite directions (larger vs. smaller). In both cases, a value close to 1 indicates that the model’s performance is nearly on par with SOTA. This approach allows for an intuitive and consistent representation of a relative comparison of the model’s performance against the best-known performance (SOTA) in a range of 0 to 1, where 1 represents perfect or near-perfect performance on the task. Note that in our analysis, the normalized value of ASR task is the average of two values estimated on LibriSpeech test-clean set and test-other set, respectively.

\begin{table*}[]
\centering
\renewcommand{\arraystretch}{1.2}
\newcolumntype{S}{>{\small}c}
\caption{Performance comparison of representative models across four core speech understanding tasks: Automatic Speech Recognition (ASR), Speech Translation (ST), Emotion Recognition (ER), and Human Sound Event Classification (SEC). The table reports both raw scores and each model's relative performance compared to the task-specific SOTA (RPS). The final two columns summarize each model’s average RPS and its standard deviation across all tasks it participates in. Since SALMONN, Kimi-Audio, and OSUM are not capable of all the tasks we tested, the asterisk * in the result indicates that the mean and standard deviation of SOTA score are calculated only based on the tasks they can handle. }

\label{table:performance}
\begin{tabular}{l|SSSSSSSS|S|S}
\hline
\multicolumn{1}{c|}{\multirow{4}{*}{\textbf{Model}}} & \multicolumn{8}{c|}{\textbf{Task-wise Performance}} & \multicolumn{1}{c|}{\multirow{5}{*}{\textbf{Avg. RPS}}} & \multicolumn{1}{c}{\multirow{5}{*}{\textbf{Std. RPS}}} \\
\cline{2-9}
\multicolumn{1}{c|}{} 
& \multicolumn{2}{c|}{ASR} 
& \multicolumn{2}{c|}{Speech Translation} 
& \multicolumn{2}{c|}{Emotion Recog.} 
& \multicolumn{2}{c|}{Sound Event Cls.} 
& \multicolumn{1}{c|}{} 
& \multicolumn{1}{c}{} \\
\cline{2-9}
\multicolumn{1}{c|}{} 
& \multicolumn{2}{c|}{LibriSpeech (clean\textbar other)} 
& \multicolumn{2}{c|}{CoVoST2 En→Zh} 
& \multicolumn{2}{c|}{MELD} 
& \multicolumn{2}{c|}{VocalSound} 
& \multicolumn{1}{c|}{} 
& \multicolumn{1}{c}{} \\
\cline{2-9}
\multicolumn{1}{c|}{} 
& \multicolumn{1}{c|}{WER $\downarrow$} & \multicolumn{1}{c|}{RPS} 
& \multicolumn{1}{c|}{BLEU $\uparrow$} & \multicolumn{1}{c|}{RPS} 
& \multicolumn{1}{c|}{ACC(\%) $\uparrow$} & \multicolumn{1}{c|}{RPS} 
& \multicolumn{1}{c|}{ACC(\%) $\uparrow$} & \multicolumn{1}{c|}{RPS} 
& \multicolumn{1}{c|}{} & \multicolumn{1}{c}{} \\
% \hline
\cline{1-9}
\textbf{SOTA} & \multicolumn{1}{c|}{1.2\textbar2.4} & \multicolumn{1}{c|}{1.00} & \multicolumn{1}{c|}{45.2} & \multicolumn{1}{c|}{1.00} & \multicolumn{1}{c|}{69.9} & \multicolumn{1}{c|}{1.00} & \multicolumn{1}{c|}{98.0} & 1.00 &  &  \\ 
\hline
\textbf{Specialized Model} & \multicolumn{1}{c|}{} & \multicolumn{1}{c|}{} & \multicolumn{1}{c|}{} & \multicolumn{1}{c|}{} & \multicolumn{1}{c|}{} & \multicolumn{1}{c|}{} & \multicolumn{1}{c|}{} &  &  &  \\
SAMBA ASR\cite{shakhadri2025samba} & \multicolumn{1}{c|}{1.2\textbar2.5} & \multicolumn{1}{c|}{0.98} & \multicolumn{1}{c|}{/} & \multicolumn{1}{c|}{/} & \multicolumn{1}{c|}{/} & \multicolumn{1}{c|}{/} & \multicolumn{1}{c|}{/} & / &  &  \\
% FAdam & \multicolumn{1}{c|}{1.3\textbar2.5} & \multicolumn{1}{c|}{0.94} & \multicolumn{1}{c|}{/} & \multicolumn{1}{c|}{/} & \multicolumn{1}{c|}{/} & \multicolumn{1}{c|}{/} & \multicolumn{1}{c|}{/} & / &  &  \\
% Parakeet-rnnt & \multicolumn{1}{c|}{1.5\textbar2.5} & \multicolumn{1}{c|}{0.88} & \multicolumn{1}{c|}{/} & \multicolumn{1}{c|}{/} & \multicolumn{1}{c|}{/} & \multicolumn{1}{c|}{/} & \multicolumn{1}{c|}{/} & / &  &  \\
% SeamlessM4T-Large & \multicolumn{1}{c|}{/} & \multicolumn{1}{c|}{/} & \multicolumn{1}{c|}{34.6} & \multicolumn{1}{c|}{0.77} & \multicolumn{1}{c|}{/} & \multicolumn{1}{c|}{/}  & \multicolumn{1}{c|}{/} & / &  &  \\
GenTranslate-V2\cite{hu2024gentranslate} & \multicolumn{1}{c|}{/} & \multicolumn{1}{c|}{/} & \multicolumn{1}{c|}{43.6} & \multicolumn{1}{c|}{0.96} & \multicolumn{1}{c|}{/} & \multicolumn{1}{c|}{/} & \multicolumn{1}{c|}{/} & / & / & / \\
ELR-GNN\cite{shou2024efficient} & \multicolumn{1}{c|}{/} & \multicolumn{1}{c|}{/} & \multicolumn{1}{c|}{/} & \multicolumn{1}{c|}{/} & \multicolumn{1}{c|}{69.9} & \multicolumn{1}{c|}{1.00} & \multicolumn{1}{c|}{/} & / &  &  \\
% BiosERC & \multicolumn{1}{c|}{/} & \multicolumn{1}{c|}{/} &  \multicolumn{1}{c|}{/} & \multicolumn{1}{c|}{/} & \multicolumn{1}{c|}{69.8} & \multicolumn{1}{c|}{1.00} & \multicolumn{1}{c|}{/} & / &  &  \\
% CKERC & \multicolumn{1}{c|}{/} & \multicolumn{1}{c|}{/} &  \multicolumn{1}{c|}{/} & \multicolumn{1}{c|}{/} & \multicolumn{1}{c|}{69.3} & \multicolumn{1}{c|}{0.99} &\multicolumn{1}{c|}{/} & / &  &  \\
% EfficientNet & \multicolumn{1}{c|}{/} & \multicolumn{1}{c|}{/} & \multicolumn{1}{c|}{/} & \multicolumn{1}{c|}{/} & \multicolumn{1}{c|}{/} & \multicolumn{1}{c|}{/} & \multicolumn{1}{c|}{90.5} & 0.92 &  &  \\
CLAP\cite{elizalde2023clap} & \multicolumn{1}{c|}{/} & \multicolumn{1}{c|}{/} & \multicolumn{1}{c|}{/} & \multicolumn{1}{c|}{/} & \multicolumn{1}{c|}{/} & \multicolumn{1}{c|}{/} & \multicolumn{1}{c|}{98.0} & 1.00 &  &  \\ \hline

% \textbf{E2E Specialized Model} & \multicolumn{1}{c|}{} & \multicolumn{1}{c|}{} & \multicolumn{1}{c|}{} & \multicolumn{1}{c|}{} & \multicolumn{1}{c|}{} & \multicolumn{1}{c|}{} & \multicolumn{1}{c|}{} &  &  &  \\
% Hubert & \multicolumn{1}{c|}{1.9\textbar3.5} & \multicolumn{1}{c|}{0.66} & \multicolumn{1}{c|}{} & \multicolumn{1}{c|}{} & \multicolumn{1}{c|}{} & \multicolumn{1}{c|}{} & \multicolumn{1}{c|}{} &  &  &  \\
% Conformer & \multicolumn{1}{c|}{1.9\textbar3.9} & \multicolumn{1}{c|}{0.62} & \multicolumn{1}{c|}{} & \multicolumn{1}{c|}{} & \multicolumn{1}{c|}{} & \multicolumn{1}{c|}{} & \multicolumn{1}{c|}{} &  &  &  \\
% Whisper-Large-v3 & \multicolumn{1}{c|}{1.8\textbar3.6} & \multicolumn{1}{c|}{0.68} & \multicolumn{1}{c|}{} & \multicolumn{1}{c|}{} & \multicolumn{1}{c|}{} & \multicolumn{1}{c|}{} & \multicolumn{1}{c|}{} &  &  &  \\ \hline
\textbf{Speech   LLM} & \multicolumn{1}{c|}{} & \multicolumn{1}{c|}{} & \multicolumn{1}{c|}{} & \multicolumn{1}{c|}{} & \multicolumn{1}{c|}{} & \multicolumn{1}{c|}{} & \multicolumn{1}{c|}{} &  &  &  \\
SALMONN & \multicolumn{1}{c|}{2.1\textbar4.9} & \multicolumn{1}{c|}{0.53} & \multicolumn{1}{c|}{33.1} & \multicolumn{1}{c|}{0.73} & \multicolumn{1}{c|}{39.2} & \multicolumn{1}{c|}{0.56} & \multicolumn{1}{c|}{/} & / & \multicolumn{1}{c|}{0.61*} & \multicolumn{1}{c}{0.11*} \\
Qwen2.5-Omni & \multicolumn{1}{c|}{2.4\textbar4.2} & \multicolumn{1}{c|}{0.54} & \multicolumn{1}{c|}{41.4} & \multicolumn{1}{c|}{0.92} & \multicolumn{1}{c|}{57.0} & \multicolumn{1}{c|}{0.81} & \multicolumn{1}{c|}{93.7} & 0.96 & \multicolumn{1}{c|}{0.81} & \multicolumn{1}{c}{0.19} \\
Qwen-Audio & \multicolumn{1}{c|}{2.0\textbar4.2} & \multicolumn{1}{c|}{0.59} & \multicolumn{1}{c|}{41.5} & \multicolumn{1}{c|}{0.92} & \multicolumn{1}{c|}{55.3} & \multicolumn{1}{c|}{0.79} & \multicolumn{1}{c|}{92.9} & 0.95 & \multicolumn{1}{c|}{0.81} & \multicolumn{1}{c}{0.17} \\
Qwen2-Audio & \multicolumn{1}{c|}{1.7\textbar4.0} & \multicolumn{1}{c|}{0.65} & \multicolumn{1}{c|}{45.2} & \multicolumn{1}{c|}{1.00} & \multicolumn{1}{c|}{55.7} & \multicolumn{1}{c|}{0.79} & \multicolumn{1}{c|}{93.8} & 0.96 & \multicolumn{1}{c|}{0.85} & \multicolumn{1}{c}{0.16} \\
Kimi-Audio & \multicolumn{1}{c|}{1.3\textbar2.4} & \multicolumn{1}{c|}{0.96} & \multicolumn{1}{c|}{/} & \multicolumn{1}{c|}{/} & \multicolumn{1}{c|}{59.1} & \multicolumn{1}{c|}{0.84} & \multicolumn{1}{c|}{94.9} & 0.97 & \multicolumn{1}{c|}{0.92*} & \multicolumn{1}{c}{0.07*} \\
Baichuan-Audio & \multicolumn{1}{c|}{3.0\textbar6.0} & \multicolumn{1}{c|}{0.40} & \multicolumn{1}{c|}{30.4} & \multicolumn{1}{c|}{0.67} & \multicolumn{1}{c|}{23.6} & \multicolumn{1}{c|}{0.34} & \multicolumn{1}{c|}{58.2} & 0.59 & \multicolumn{1}{c|}{0.50} & \multicolumn{1}{c}{0.16} \\
OSUM & \multicolumn{1}{c|}{2.2\textbar5.5} & \multicolumn{1}{c|}{0.49} & \multicolumn{1}{c|}{/} & \multicolumn{1}{c|}{/} & \multicolumn{1}{c|}{53.4} & \multicolumn{1}{c|}{0.76} & \multicolumn{1}{c|}{82.6} & 0.84 & \multicolumn{1}{c|}{0.70*} & \multicolumn{1}{c}{0.18*} \\ \hline
\end{tabular}
\end{table*}

\subsubsection{Performance Comparison and Key Insights}
In this section, we analyze the performance of various models across four core speech understanding tasks: ASR, ST, Emotion Recognition (ER), and Human Sound Event Classification (SEC). We focus on comparing Speech LLMs to both specialized models and the state-of-the-art (SOTA) in each task. Table \ref{table:performance} summarizes the results of the model performance across these tasks.

\paragraph{Competitive Performance of Speech LLMs in Selected Tasks} The table clearly illustrates that top-tier Speech LLMs demonstrate competitive performance across tasks from all three categories. For example, in Automatic Speech Recognition (ASR), Kimi-Audio outperforms several specialized models, achieving a Word Error Rate (WER) of 2.4 on the LibriSpeech test-other set, which matches the current SOTA performance. Most other Speech LLMs achieve a WER around 2.0 on the LibriSpeech test-clean set and approximately 4.0 on the test-other set, which, although slightly below SOTA, still represent commendably high performance in ASR. In the domain of Speech Translation (ST), Qwen2-Audio achieves SOTA performance on the CoVoST2 En→Zh task, establishing itself as one of the best-performing models for speech translation. This result highlights that Speech LLMs are not confined to purely linguistic tasks such as ASR, but also excel in translating spoken language into written text, showcasing their versatility across tasks.

For the Emotion Recognition (ER) task, both Qwen2.5-Omni and Kimi-Audio demonstrate competitive performance, with RPS surpassing 0.8. While these scores are lower than those achieved by specialized models, they nevertheless indicate that Speech LLMs can still effectively perform emotion recognition tasks, thereby extending their applicability to nuanced paralinguistic domains. Similarly, Speech LLMs perform strongly in Human Sound Event Classification, with models like Qwen2-Audio and Kimi-Audio achieving RPS of 0.96 and 0.97, respectively. These results underscore the ability of Speech LLMs to recognize and classify sound events with an accuracy comparable to top-tier specialized models. The strong performance of Speech LLMs in these paralinguistic and non-linguistic tasks is particularly significant, considering that most Speech LLMs are based on encoders that are primarily trained for ASR. Their ability to generalize effectively to tasks beyond the scope of traditional speech recognition demonstrates the considerable potential of Speech LLMs. This suggests that their capabilities are not confined to a narrow set of speech tasks, but extend across a broader range of complex perception and cognition tasks.

\paragraph{Multitask Capabilities of Speech LLMs} The table also highlights the impressive multitask capabilities of Speech LLMs, as evidenced by their consistent performance across multiple tasks. Notably, Kimi-Audio achieves an average RPS of 0.92, reflecting its robust multitask performance across its evaluated tasks. This performance is particularly notable given that Kimi-Audio excels not only in ASR but also in paralinguistic tasks and non-linguistic tasks, demonstrating a high level of versatility. Similarly, Qwen-Audio series models, such as Qwen-Audio and Qwen2-Audio, achieve average RPS of 0.81 and 0.85, respectively. These results reinforce that Speech LLMs can effectively handle diverse speech-related tasks, spanning tasks focusing on different information in the audio. 

Even more strikingly, Qwen2.5-Omni, a model designed to handle multimodal tasks beyond speech, still achieves a competitive average RPS of 0.81 across the speech tasks. This demonstrates that Qwen2.5-Omni’s performance in speech tasks is not compromised by its broader multimodal capabilities. In fact, its performance remains comparable with other Speech LLMs, reinforcing the idea that Speech LLMs can successfully scale to support multiple modalities without losing efficacy in speech-specific tasks. This highlights the generalizability of Speech LLMs to multimodal LLMs, showing that their core speech understanding abilities can be preserved and even enhanced when extended to other modalities, further underscoring their adaptability for future cross-domain applications.

These observations highlight the growing multi-task capabilities of Speech LLMs. This underscores the generalizability of Speech LLMs across various domains, demonstrating their ability to perform well not only in specific tasks but also across a broad range of applications, marking a significant advancement in speech understanding.

\begin{table}[htbp]
  \small
  \renewcommand{\arraystretch}{1.25}
  \centering
  \caption{Performance comparison of representative models on typical reasoning benchmarks. Score refers to the score of LLM judge.}
  \label{tab:model_performance}
  
  \begin{tabular}{p{2cm} p{4.2cm} p{1.5cm}}
    \toprule
    \textbf{BenchMark} & \textbf{Top Models} & \textbf{Acc / Score} \\
    \midrule
    \multirow{11}{*}{MMAR} 
    & Gemini 2.0 Flash & 65.60 \\
    & GPT-4o Audio & 63.50 \\
    & Qwen-2.5-Omni (7B)~\cite{xu2025qwen2} & 56.70 \\
    & Qwen-2.5-Omni (3B)~\cite{xu2025qwen2} & 53.80 \\
    & Baichuan-Omni-1.5 (11B)~\cite{li2025baichuan} & 40.70 \\
    & Audio-Reasoner (8.4B)~\cite{xie2025audio} & 36.80 \\
    & SALMONN (13B)~\cite{tang2023salmonn} & 33.20 \\
    & SALMONN (7B)~\cite{tang2023salmonn} & 32.80 \\
    & Audio-CoT (8.4B)~\cite{ma2025audio} & 31.30 \\
    & Qwen2-Audio (7B)~\cite{chu2024qwen2} & 30.40 \\
    & Qwen2-Audio-Instruct (7B)~\cite{chu2024qwen2} & 30.00 \\
    \midrule

    \multirow{5}{*}{MMAU} 
    & Gemini 2.0 Flash & 59.93 \\
    & Gemini Pro v1.5 & 52.97 \\
    & Qwen2-Audio-Instruct (7B)~\cite{chu2024qwen2} & 52.5 \\
    & Qwen2-Audio-Chat (7B)~\cite{chu2024qwen2} & 41.86 \\
    & SALMONN (13B)~\cite{tang2023salmonn} & 32.77 \\
    \midrule

    \multirow{3}{*}{AIR-Bench} 
    & Qwen-Audio-Turbo (7B)~\cite{chu2023qwen} & 57.8 / 6.34 \\
    & Qwen-Audio-Chat (7B)~\cite{chu2023qwen} & 54.5 / 6.08 \\
    & SALMONN~\cite{tang2023salmonn} & 36.0 / 6.11 \\
    \bottomrule
  \end{tabular}
\end{table}

\paragraph{Variability in Performance Across Tasks} Despite the promising multitask capabilities of Speech LLMs, there is significant variability in their performance across different tasks. Task-wise speaking, although leading models consistently achieve RPS greater than 0.9 in both Speech Translation and Human Sound Event Classification, their performance in Emotion Recognition (ER) is less impressive, with RPS falling to around 0.8. This indicates that, while Speech LLMs can perform admirably in certain domains, their generalization to emotion-related tasks is still a challenge.

Additionally, examining the standard deviation of RPS provides further insight into the performance variability. Models like the Qwen-Audio series exhibit a standard deviation of 0.15–0.20 in their RPS across tasks, suggesting that their performance can be inconsistent across different domains. While these models are strong contenders in many areas, this variability underscores the limited multi-task ability of Speech LLMs, as they do not always maintain high performance across all tasks. This inconsistency reflects the challenges that still exist in making Speech LLMs robust across diverse tasks.

Furthermore, an important aspect to note is that not all models are capable of performing all the tasks tested. For example, SALMONN, Kimi-Audio, and OSUM do not report performance on all tasks in the table. For the tasks they do not report, we conducted our own experiments. We found that SALMONN performs poorly in Human Sound Event Classification, with accuracy close to random guessing. Kimi-Audio and OSUM, on the other hand, are not capable of Speech Translation and consistently output transcriptions or nonsensical text when tested. For consistency, the mean and standard deviation of RPS in the table are calculated only on the tasks these models can handle. Although Kimi-Audio exhibits consistently strong performance on the tasks it has been trained for—reflected in its high average RPS and low standard deviation—it fails to handle Speech Translation, a task outside its training scope. This limitation underscores a broader challenge in current Speech LLMs: while they can generalize well within familiar domains, their performance does not automatically transfer to tasks for which they lack explicit training. Despite sharing similarities with other related tasks, this capability does not automatically generalize, underscoring the current limitations in the flexibility and generalization of Speech LLMs.

\paragraph{ Limited Performance on Deep Understanding Tasks} 

While the previous section provides a systematic analysis of current Speech LLM capabilities from the informational perspective, it primarily focuses on tasks mainly involving perception and shallow cognition. From the standpoint of cognitive depth, as illustrated in Section \ref{taxonomy:functional}, however, it offers limited evaluation of tasks requiring deeper reasoning or complex inference. To address this gap, several recent studies have proposed dedicated benchmarks targeting such cognitively demanding tasks and have conducted comprehensive evaluations of leading Speech LLMs. The results of these evaluations are summarized in Table \ref{tab:model_performance}. These benchmarks evaluate models on closed-ended or mixed open/closed-ended question answering tasks.

In addition to tasks that primarily capture acoustic content or straightforward semantic inference, these benchmarks probe cognitively richer abilities, such as emotion-state summarization from speech and multi-speaker role mapping. For example, in the MMAU benchmark, the Speech Emotion State Summarisation and Multi-Speaker Role Mapping tracks evaluate how well models integrate prosody, timbre, and speaker-dependent cues with textual reasoning. These tasks highlight that Speech LLMs must jointly model fine-grained acoustic signals (e.g., emotional valence, speaker traits, turn-taking structure) and textual semantics to achieve genuine speech cognition. 

As shown in Table~\ref{tab:model_performance}, current Speech LLMs still exhibit suboptimal performance on benchmarks involving deeper levels of speech understanding. While proprietary models like Gemini 2.0 Flash and GPT-4o Audio lead with accuracy scores around 60--66\%, most open-source systems perform significantly worse, with many results falling below 35\%~\cite{ma2025mmar}. This \textit{performance gap} underscores the difficulty of deriving high-level semantic inferences from spoken input. Compared to perception or shallow cognitive tasks, these benchmarks require contextual reasoning, discourse-level understanding, and multimodal integration—areas where current models remain limited in capability.

The observed variability, lack of robust multitask generalization, and underperformance on deep understanding tasks highlight the critical challenges facing current Speech LLMs. In the next chapter, we will explore these challenges in depth, focusing on issues like LLM dormancy, the limitations in semantic reasoning abilities, and the inadequate capture and reasoning of acoustic information. Addressing these challenges is critical for further advancing Speech LLMs and expanding their applicability across a broader range of speech understanding tasks.

\subsection{Model Performance in Real-World Scenarios}
\label{sec:real_world_performance}

While the previous subsections evaluate Speech LLMs on standard benchmarks, real-world deployment introduces additional considerations beyond those captured in curated datasets. In practice, user speech often contains background noise, interruptions, informal phrasing, and diverse recording conditions. As a result, performance in in-the-wild scenarios may deviate from results obtained on clean academic benchmarks. For example, several recent Speech LLMs report stable performance on large-scale real-world corpora such as WenetSpeech~\cite{zhang2022wenetspeech} and GigaSpeech~\cite{chen2021gigaspeech}, suggesting that their robust acoustic front-ends and large-scale pretraining help maintain recognition quality in unconstrained settings. However, the lack of a unified and widely accepted real-world evaluation benchmark makes it difficult to directly compare models across practical application contexts.

In addition, real-world systems often need to satisfy specific deployment constraints. One such challenge arises in \textit{low-resource} and \textit{multilingual} scenarios, where available speech data is sparse or covers diverse accents and dialects. Current Speech LLMs often rely on high-resource pretraining and thus do not consistently outperform traditional ASR pipelines or multilingual speech models in these settings, as shown in Table \ref{tab:model_performance_multilingual}~\cite{xue2024ideal, li2025mosamixturessimpleadapters, fang2025lowresourcedomainadaptationspeech}. Another challenge is \textit{real-time interaction}. The large parameter scales of many Speech LLMs lead to high inference latency, posing difficulties for latency-sensitive applications such as interactive voice assistants, customer service agents, and dialogue-driven human–robot interaction. Although quantization, model distillation, and streaming decoding architectures have been explored to reduce latency, real-time, fully multimodal LLM-based interaction remains an active research problem.

Overall, while Speech LLMs demonstrate increasingly competitive robustness in both controlled evaluation settings and real-world corpora, their performance and practicality in truly diverse, noise-rich, multilingual, and low-latency scenarios remain limited. Addressing these challenges is essential for advancing Speech LLMs toward reliable, real-world conversational intelligence.

\begin{table}[htbp]
  \small
  \renewcommand{\arraystretch}{1.25}
  \centering
  \caption{Comparison of multilingual ASR performance on the FLEURS dataset (average WER across 19 languages).\label{tab:model_performance_multilingual}}
  
  \begin{threeparttable}
  \begin{tabular}{p{2.7cm} p{4.2cm} p{1.0cm}}
    \toprule
    \textbf{Category} & \textbf{Model} & \textbf{WER $\downarrow$} \\
    \midrule

    \multirow{2}{*}{E2E specialized model} 
      & Whisper Large-v2 & 8.86 \\
      & Whisper Large-v3 & 6.52 \\
    \midrule

    \multirow{3}{*}{Open-source SLLM} 
      & Qwen2.5-Omni & 14.04 \\
      & Qwen3-Omni-30B-A3B-Instruct & 5.33 \\
      & Qwen3-Omni-Flash-Instruct & 5.31 \\
    \midrule

    \multirow{2}{*}{Proprietary SLLM} 
      & GPT-4o-Transcribe & 4.48 \\
      & Gemini-2.5-Pro & 5.55 \\
    \bottomrule
  \end{tabular}

  \begin{tablenotes}
    \footnotesize
    \item \textit{Languages included}: Arabic, Cantonese, Chinese, Dutch, English, French, German, Indonesian, Italian, Japanese, Korean, Malay, Portuguese, Russian, Spanish, Thai, Turkish, Urdu, and Vietnamese.
  \end{tablenotes}
  \end{threeparttable}

\end{table}

\section{Challenges}
\label{sec:challenge}

To further investigate the limitations observed in the performance section, we conducted a series of experiments aimed at clarifying and localizing the underlying issues. Building on the multidimensional perspective introduced in Section \ref{taxonomy:functional}, we identify two main categories of challenges.

\subsection{Instruction sensitivity} 
\label{sec: instruction_sensitivity}
\label{sec:instr_sensitivity}
\begin{table*}[t]
\centering
\caption{Sensitivity of different instruction variations for ASR, S2TT, SER and GR tasks among three Speech LLMs. The test sets for the four tasks are: LibriSpeech~\cite{panayotov2015librispeech} (test-clean / test-other) for ASR, LibriSpeech (test-clean) for GR, IEMOCAP~\cite{busso2008iemocap} (session 5) for SER, and CoVoST2~\cite{wang2021covost} (en2zh test set) for S2TT.}
\label{tab:sensitivity}
\renewcommand{\arraystretch}{1.3}
\small
\setlength{\tabcolsep}{3pt} 
\resizebox{0.9\textwidth}{!}{
\begin{tabular}{c|c|c|ccccccc c}
\hline
\multirow{3}{*}{\textbf{Model}} &
\multirow{3}{*}{\textbf{Task}} &
\multirow{3}{*}{\textbf{Metrics}} &
\multicolumn{7}{c}{\textbf{Instruction Variation}} &
\multirow{2}{*}{\textbf{Sensitivity}} \\
\cline{4-10}
& & &
\multirow{2}{*}{\textbf{Default}} &
\multirow{2}{*}{\textbf{Punctuation}} &
\multicolumn{3}{c}{\textbf{Semantic Complexity}} &
\multicolumn{2}{c}{\textbf{Case}} & \\
\cline{6-8}\cline{9-10}\cline{11-11}
& & & & &
\textbf{simple} & \textbf{neutral} & \textbf{complex} &
\textbf{lower} & \textbf{upper} &
\textbf{std.} \\
\hline
% ───────────────────────── Qwen2-Audio-Instruct-7B ─────────────────────────
\multirow{8}{*}{\begin{tabular}[c]{@{}c@{}}Qwen2-Audio\\-7B-Instruct\end{tabular}}
% --------------- ASR ---------------
& \multirow{2}{*}{\begin{tabular}[c]{@{}c@{}}ASR\end{tabular}} & IFR
& 98.59/98.67 & 95.53/94.32 & 94.73/93.84 & 83.59/80.91 & 29.01/25.52 & 98.47/98.33 & \textbf{99.20/99.25} & \multirow{2}{*}{22.78} \\
& & WER$_{\text{IF}}$(\%)$\downarrow$
& 4.89/7.79 & 8.86/13.85 & 10.30/14.28 & 22.21/28.34 & 72.20/76.89 & 5.33/8.53 & \textbf{3.75/6.89} & \\ \cline{2-11}
% --------------- S2TT ---------------
& \multirow{2}{*}{\begin{tabular}[c]{@{}c@{}}S2TT\end{tabular}} & IFR
& 95.10 & 92.79 & 63.14 & \textbf{95.90} & 94.55 & 92.33 & 91.70 & \multirow{2}{*}{5.68} \\
& & BLEU$_{\text{IF}}$$\uparrow$
& \textbf{37.76} & 36.68 & 20.81 & 37.72 & 36.98 & 36.63 & 35.96 & \\ \cline{2-11}
% --------------- SER ---------------
& \multirow{2}{*}{SER} & IFR
& \textbf{100} & \textbf{100} & \textbf{100} & \textbf{100} & \textbf{100} & \textbf{100} & \textbf{100} & \multirow{2}{*}{1.25} \\
& & ACC$_{\text{IF}}$(\%)$\uparrow$
& 65.51 & 65.43 & 64.14 & \textbf{66.40} & 63.26 & 65.11 & 62.61 & \\ \cline{2-11}
% --------------- GR ---------------
& \multirow{2}{*}{GR} & IFR
& \textbf{100} & \textbf{100} & \textbf{100} & \textbf{100} & \textbf{100} & \textbf{100} & \textbf{100} & \multirow{2}{*}{0.81} \\
& & ACC$_{\text{IF}}$(\%)$\uparrow$
& 89.12 & 88.85 & 88.28 & 88.78 & \textbf{91.03} & 89.05 & 89.43 & \\
\hline
% ───────────────────────── Qwen2.5-Omni-7B (占位) ─────────────────────────
\multirow{8}{*}{\begin{tabular}[c]{@{}c@{}}Qwen2.5-Omni\\-7B\end{tabular}}
% ----- ASR -----
& \multirow{2}{*}{\begin{tabular}[c]{@{}c@{}}ASR\end{tabular}} & IFR
& 98.21/96.73 & 98.32/97.11 & 98.66/97.38 & 95.80/94.15 & 98.63/97.28 & \textbf{99.01/97.52} & 68.70/69.82 & \multirow{2}{*}{15.48} \\
& & WER$_{\text{IF}}$(\%)$\downarrow$
& 4.61/8.68 & 4.48/8.14 & 3.77/7.49 & 5.64/9.37 & \textbf{3.27/6.92} & 3.92/7.92 & 48.48/47.97 & \\ \cline{2-11}
% ----- S2TT -----
& \multirow{2}{*}{\begin{tabular}[c]{@{}c@{}}S2TT\end{tabular}} & IFR
& 99.48 & 99.42 & 95.54 & \textbf{99.61} & 94.56 & 99.43 & 98.69 & \multirow{2}{*}{1.02} \\
& & BLEU$_{\text{IF}}$$\uparrow$
& \textbf{45.28} & 45.22 & 42.46 & 45.01 & 43.18 & 44.49 & 43.64 & \\ \cline{2-11}
% ----- SER -----
& \multirow{2}{*}{SER} & IFR
& 93.71 & 94.84 & 96.62 & \textbf{98.39} & 97.67 & 94.76 & 26.91 & \multirow{2}{*}{14.66} \\
& & ACC$_{\text{IF}}$(\%)$\uparrow$
& 58.90 & 59.63 & 53.02 & \textbf{64.95} & 57.37 & 58.58 & 17.89 & \\ \cline{2-11}
% ----- GR -----
& \multirow{2}{*}{GR} & IFR
& 68.85 & 75.27 & \textbf{77.06} & 54.77 & 14.27 & 72.98 & 2.94 & \multirow{2}{*}{20.69} \\
& & ACC$_{\text{IF}}$(\%)$\uparrow$
& 51.87 & \textbf{56.98} & 49.58 & 40.50 & 11.37 & 55.53 & 2.10 & \\
\hline
% ───────────────────────── SALMONN-13B (占位) ─────────────────────────
\multirow{8}{*}{\begin{tabular}[c]{@{}c@{}}SALMONN\\-13B\end{tabular}}
% ----- ASR -----
& \multirow{2}{*}{\begin{tabular}[c]{@{}c@{}}ASR\end{tabular}} & IFR
& 99.92/99.69 & \textbf{99.92/99.73} & 99.73/99.52 & 99.92/99.69 & 99.73/99.76 & 99.92/99.69 & 96.79/97.45 & \multirow{2}{*}{0.86}\\
& & WER$_{\text{IF}}$(\%)$\downarrow$
& 2.21/5.27 & \textbf{2.22/5.15} & 2.58/5.98 & 2.22/5.29 & 2.35/5.39 & 2.18/5.26 & 4.72/7.15 & \\ \cline{2-11}
% ----- S2TT -----
& \multirow{2}{*}{\begin{tabular}[c]{@{}c@{}}S2TT\end{tabular}} & IFR
& 99.94 & 99.95 & 99.67 & 99.95 & 99.94 & \textbf{99.96} & 99.93 & \multirow{2}{*}{0.21}\\
& & BLEU$_{\text{IF}}$$\uparrow$
& 34.53 & 34.51 & 34.36 & 34.41 & 33.87 & \textbf{34.49} & 34.35 & \\ \cline{2-11}
% ----- SER -----
& \multirow{2}{*}{SER} & IFR
& \textbf{100} & \textbf{100} & \textbf{100} & \textbf{100} & \textbf{100} & \textbf{100} & \textbf{100} & \multirow{2}{*}{3.20}\\
& & ACC$_{\text{IF}}$(\%)$\uparrow$
& 39.08 & 39.73 & 32.39 & \textbf{42.47} & 34.09 & 39.08 & 37.63 & \\ \cline{2-11}
% ----- GR -----
& \multirow{2}{*}{GR} & IFR
& \textbf{100} & \textbf{100} & 97.98 & \textbf{100} & \textbf{100} & \textbf{100} & \textbf{100} & \multirow{2}{*}{10.20}\\
& & ACC$_{\text{IF}}$(\%)$\uparrow$
& 47.10 & 47.02 & \textbf{76.53} & 47.18 & 48.32 & 47.29 & 47.56 & \\
\hline
\end{tabular}
}
\end{table*}

% From the perspective of perceptual and shallow cognitive tasks, we highlight the model's sensitivity to instruction adherence—an issue we refer to as LLM Dormancy. This phenomenon is examined through tasks such as automatic speech recognition (ASR) and speech translation (ST), which are characterized by well-defined objectives and simple, precise prompts. The core requirement of these tasks is reliability.

Instruction sensitivity in LLMs has been extensively explored within natural language processing (NLP), where it is widely regarded as a key indicator of model robustness~\cite{liu2023lost, sclar2024quantifying}. Robustness refers to the ability of LLMs to maintain consistent task performance when faced with semantically equivalent but linguistically varied instructions. These variations may include character-level differences (e.g., punctuation), synonym substitution, rephrased sentence structure, or even different requested output formats.

In the audio domain, the IFEval-Audio benchmark~\cite{xu2024ifeval} provides a six-dimensional perspective on instruction-following performance in end-to-end or cascaded LLMs that integrate the audio modality. It evaluates models along six dimensions: Content, Capitalization, Symbol, List Structure, Length, and Format. However, this benchmark only assesses dialog performance, focusing on whether models preserve the semantics of chat and QA interactions, and uses LLMs to verify if responses maintain semantic consistency. Although useful, this is far from sufficient for the unified speech understanding LLMs studied in this work, which require evaluation across a broader range of SLU tasks beyond conversational compliance.

Speech LLMs are expected to produce reliable outputs for SLU tasks, which require both acoustic perception and shallow semantic reasoning. Although these models typically perform well on instructions encountered during training, two critical challenges emerge in real-world scenarios where the same or similar task requirements are expressed using varied instructions: 
\begin{enumerate}
    \item \textbf{Instruction following failure:} the model sometimes fails to correctly follow the given instruction.
    \item \textbf{Instruction-induced performance instability:} Even when the model appears to follow the instruction, its performance on the task may fluctuate unpredictably.
\end{enumerate}

These issues pose significant practical limitations. For example, in batch-processing pipelines, variation in instruction can lead to operational inefficiencies and reduced reliability. To mitigate this, practitioners often resort to manual instruction selection—testing multiple phrasings and selecting the one that yields the best performance for each task. However,  this approach is labor-intensive and inherently unstable. To achieve more consistent performance, practitioners frequently fine-tune models using parameter-efficient fine-tuning methods (PEFT) ~\cite{hu2021lora, dettmers2023peft}. Although effective in stabilizing outputs, this fine-tuning typically compromises the model’s generalization capabilities, resulting in overfitting to a narrow instruction set. Additionally, PEFT introduces computational overhead and increases deployment time, making the deployment pipeline more resource-intensive.

In this survey, we formally define these phenomena as key challenges for speech LLMs in speech understanding tasks. As a preliminary step, we conduct empirical evaluations to quantify their effects. To assess acoustic perception sensitivity, we employ automatic speech recognition (ASR) and gender recognition (GR) tasks. To evaluate shallow cognition sensitivity, we utilize speech-to-text translation (S2TT) and speech emotion recognition (SER).

\paragraph{\textbf{Instruction following measurement}}
To comprehensively evaluate the sensitivity of speech LLMs for SLU tasks, we first introduce instruction following rate(IFR) to assess the model’s ability to handle the challenge of instruction following failure. This metric reflects the proportion of samples in the evaluation set for which the model is able to reasonably follow the given instructions. In our experiments, we adopted concise and explicit instructions which instruct the model to return only the task-specific answer, omitting any extraneous or conversational content. To assess instruction-following ability, we used insertion errors in the WER calculation as an indicator. Outputs with more than 2 insertions usually contained extraneous content beyond the transcription, for example prefixes like ``The transcription is'', or even omitted the target language altogether. In such cases, the output was deemed to have failed to follow the instruction. For the S2TT task, we directly searched for specific prefixes in the model output and then conducted a secondary verification against the ground truth to determine whether the instruction was not followed. In the SER task, the follow up of the instruction was assessed by checking whether the output was strictly one of the four expected categories: ``Happy", ``Sad", ``Angry", or ``Neutral". Similarly, for the GR task, the output was required to be either ``Male" or ``Female".

\paragraph{\textbf{Performance instability measurement}}
To address the challenge of instruction-induced performance instability, we treat samples where the instruction is not followed as empty outputs and compute performance metrics accordingly. Suppose $\mathcal{D} = (\mathcal{G}, \mathcal{H})$ is the paired evaluation dataset, where $\mathcal{G} = \{g_i\}$ denotes the set of ground-truth outputs and $\mathcal{H} = \{h_i\}$ the corresponding model outputs. Let $\mathcal{F} \subseteq \mathcal{H}$ denote the subset of outputs that correctly follow the given instructions. The resulting performance under the constraint of instruction following adherence is thus referred to as

\begin{align}
    \textbf{Metric}_{\text{IF}}(\mathcal{D}) = \frac{\sum_{h_i\in \mathcal{F}}\textbf{Metric}(g_i, h_i)+\sum_{h_i\notin \mathcal{F}}\textbf{Metric}(g_i, \varnothing)}{|\mathcal{D}|},
\end{align}
i.e. the output will be set to an empty string when the instruction is not followed.

For original metrics, we adopt \textit{Word Error Rate}(WER) for the ASR task, BLEU score for the S2TT task from English to Mandarin, and classification accuracy(ACC) for both GR and SER tasks.

\paragraph{\textbf{Instruction sensitivity evaluation and analysis}}
In this dimension of analysis, we first construct a general instruction and then design six variant instructions along different perturbation axes. These variants are designed to systematically examine the model’s sensitivity to changes in instruction phrasing, including alterations in punctuation, semantic complexity, case sensitivity, and format requirements. The details of the instructions variants are demonstrated in Appendix~\ref{sec: instruction variants}. The goal is to evaluate the robustness and consistency of the instruction following in diverse transformations with consistent SLU task requirements. Three open-source models are evaluated, which are Qwen2-Audio-7B-Instruct, Qwen2.5-Omni-7B and SALMONN-14B. The instruction sensitivity of a model is positively correlated with the standard deviation of the performance instability metric set containing $\textbf{Metric}_{\text{IF}}$ on different axes, indicating that greater variability in performance across different instructions reflects higher sensitivity to instruction phrasing:
\begin{align}
    \text{Instruction Sensitivity} \propto \text{std}(\{\textbf{Metric}_{\text{IF}}\}).
\end{align}

As demonstrated in Table~\ref{tab:sensitivity}, we utilize the standard errors of $\{\textbf{Metric}_{\text{IF}}\}$ to represent instruction sensitivity. To varying degrees, it is observed across all evaluated models, which poses a potential challenge to the models’ fundamental perception and cognition capabilities. The results indicate that Qwen2.5-Omni-7B exhibits relatively high instruction sensitivity, with its Instruction-Following Rate (IFR) showing sharp declines in certain tasks. This may be partially attributed to its Omni architecture. In contrast, SALMONN-13B performs poorly on most tasks, with the exception of ASR. On the GR task, its accuracy occasionally falls below that of a random guesser with 50\% accuracy, likely because it consistently outputs the same label. Nevertheless, it demonstrates exceptionally low instruction sensitivity, maintaining an instruction-following rate close to 100\%. We hypothesize that a model’s degree of instruction sensitivity is closely related to the types of instructions it encountered during fine-tuning and the diversity of task-specific data it is trained on. 

Instruction sensitivity is a noteworthy challenge that warrants further attention. In our experiments, only the phrasing of the instructions is altered; however, it is plausible that variations in the expected output format specified by the instruction template could similarly affect model behavior. This possibility calls for deeper investigation and analysis in future research.

\subsection{Degradation on semantic understanding ability}
\label{sec:challenge2}
\begin{table*}[t]
    \centering
    % Preamble: \usepackage{multirow,makecell}
    \caption{Challenge B (\cref{sec:challenge2}): comparison of Speech LLMs and base LLMs on ST (Covost2) and SLU (MMSU). MMSU dataset does not contain GT text transcription of the speech, so the corresponding rows are emitted.}
    \renewcommand\arraystretch{1.2}
    \setlength\tabcolsep{3pt}
    \begin{tabular}{l|l|Hc|Hccccc}
    \toprule 
    %\hline
    \multirow{2}{*}{\textbf{Model}} 
    & \multirow{2}{*}{\textbf{Input}} 
    & \multicolumn{2}{c|}{\textbf{ST (Covost2)}~\cite{wang2021covost}} 
    & \multicolumn{6}{c}{\textbf{SLU (MMSU)}~\cite{wang2025mmsu}} \\ \cline{3-10}
    & & \textbf{IFR}(\%)$\uparrow$  & \textbf{BLEU}$\uparrow$ & \textbf{IFR}(\%)$\uparrow$ & \textbf{\makecell{Intent\\Detection}}(\%)$\uparrow$ 
    & \textbf{\makecell{Causal\\Reasoning}}(\%)$\uparrow$ & \textbf{\makecell{Logical\\Reasoning}}(\%)$\uparrow$
    & \textbf{\makecell{Polysemy\\Reasoning}}(\%)$\uparrow$ & \textbf{\makecell{Overall\\Acc.}}(\%)$\uparrow$ \\ \hline
    Random Baseline             & -        & -    & -    & -     & 25.00 & 25.00 & 25.00 & 25.00 & 25.00 \\ \hline
    \multirow{3}{*}{Qwen2.5-Omni~\cite{xu2025qwen2}} 
     & GT Text   & 93.86 & \textbf{45.87} &  -   & -    & -    & -    & -    & -    \\
     & ASR Text  & 94.62 & 41.98 &   100.00   & 86.09 & \textbf{100.00}& 79.81 & \textbf{96.94} & 90.76 \\
     & Speech    & 86.04 & 39.33 &   97.00  & 83.48 & 94.83 & 70.19 & 95.92 & 86.14 \\ \hline
    \multirow{2}{*}{\makecell{Qwen2.5-7B-Instruct~\cite{qwen2025qwen25technicalreport}}} 
     & GT Text   & 100.00& 43.97 &   -  & -    & -    & -    & -    & -    \\
     & ASR Text  & 100.00& 40.59 &   100.00  & \textbf{88.70} & 99.14 & \textbf{80.77} & \textbf{96.94} & \textbf{91.45} \\ \hline \hline
    \multirow{3}{*}{SALMONN~\cite{tang2023salmonn}} 
     & GT Text   & -    &  32.53  & -     & -    & -    & -    & -    & -    \\
     & ASR Text  & -    &  29.26  & 100.00 & 30.43 & 79.31 & 33.65 & 73.47 & 54.04 \\
     & Speech    & -    &  \textbf{33.50}  & 100.00 & 33.04 & 78.45 & 33.65 & 71.43 & 54.04 \\ \hline
    \multirow{2}{*}{Vicuna-13B-v1.1~\cite{zheng2023judging}} 
     & GT Text   & -    &  22.48  & -     & -    & -    & -    & -    & -    \\
     & ASR Text  & -    &  20.35  & 95.62  & \textbf{37.39} & \textbf{89.66} & \textbf{36.54} & \textbf{77.55} & \textbf{60.28} \\ \bottomrule
    \end{tabular}

    %\todo{caption}
    \label{tab:challenge-2}
\end{table*}
While Speech LLMs acquire the ability to understand speech representations, their architectures and parameter sizes remain unchanged compared to their base LLMs. This suggests a potential compromise in the original semantic and textual comprehension capabilities of the LLMs. 

To examine this hypothesis, we investigate the semantic reasoning capabilities of Speech LLMs by comparing their performance with their original base LLMs under matched semantic inputs.
To ensure fair comparison, we design a two-stream experimental setting where the base LLM receives ASR-transcribed text from the Speech LLM alongside the same instruction, while the Speech LLM receives the raw audio and the same instruction.
% we compared the performance of Speech LLMs and their corresponding base LLMs (such as SALMONN~\cite{tang2023salmonn} and Vicuna-13B-v1.1~\cite{zheng2023judging}) on speech translation and speech semantics understanding tasks. 

We evaluate two representative Speech LLMs—Qwen2.5-Omni~\cite{xu2025qwen2} and SALMONN~\cite{tang2023salmonn}—and their respective base LLMs—Qwen2.5-7B-Instruct~\cite{qwen2025qwen25technicalreport} \footnote{We choose the finetuned variant for better instruction-following.} and Vicuna-13B-v1.1~\cite{zheng2023judging}—on both ST and SLU tasks. For ST, we use the Covost2~\cite{wang2021covost} dataset with the English-to-Chinese subset. For SU, we use a curated 433 sample subset of MMSU~\cite{wang2025mmsu} containing four categories of semantic reasoning tasks: intent detection, causal reasoning, logical reasoning, and polysemy reasoning. We use the official prompt\footnote{\href{https://github.com/dingdongwang/mmsu_bench}{https://github.com/dingdongwang/mmsu\_bench}} for MMSU test. Evaluation metrics include BLEU for ST, and category-wise accuracy for SU. We do not examine IFR in this challenge, and simply consider the whole output as answer for the ST task and the first capital letter as the chosen option for the SLU task. 

Table~\ref{tab:challenge-2} shows the results. And the detailed prompts are listed in Section \ref{sec:challenge2-prompt}.
In the ST task, from the perspective of Qwen-Omni, we observe that while its overall translation performance improves compared to the Instruct variant due to better task fine-tuning, the results across the three input conditions (Speech, ASR Text, and GT Text) indicate that integrating the speech modality leads to less satisfactory end-to-end transcription results.
For the SALMONN experiments, given that Vicuna lacks sufficient instruction tuning, its translation performance is noticeably inferior to the SALMONN model. Moreover, within the SALMONN series, the integration of the speech modality demonstrates relatively strong overall performance, demonstrating the effectiveness of the Speech LLM paradigm on shallow cognitive tasks.

However, in the unseen SLU task, both Qwen2.5-Omni and SALMONN show drops compared to their base text LLMs. Additionally, both Qwen2.5-Omni and SALMONN demonstrates slightly higher or comparable performance accuracy when given ASR text as input compared to direct speech input, which implies that the benefits of two-stage step-by-step reasoning are not inferior to the potential degradation caused by error propagation in the cascaded pipeline in task. It further demonstrates the potential decline in deep semantic reasoning ability caused by multimodal fusion during the training process of current Speech LLM paradigms.

In general, these results suggest that although Speech LLMs can align speech inputs with textual outputs through fine-tuning, their original textual reasoning capacity may be weakened, likely due to modality alignment constraints~\cite{fang2025lowresourcedomainadaptationspeech} and shared model capacity.

Beyond the two challenges discussed above, Speech LLMs encounter several additional limitations. 
First, they often struggle to reliably capture and reason over acoustic information, leading to restricted performance in tasks that require fine-grained auditory understanding. 
Another challenge lies in the temporal alignment of speech segments, in particular the definition of independent time boundaries using instruction-based mechanisms. This aspect has been relatively underexplored, as it is tightly coupled with downstream tasks. 
Finally, managing speaker turns and preserving correct sequential order in multi-party conversations remain open problems that warrant further investigation.

\section{Future Exploration}
\label{sec:print}

Building on the challenges discussed in Section \ref{sec:challenge}, as Speech LLMs continue to evolve, addressing key challenges will be crucial for their progress. Two primary issues that warrant focused research are instruction sensitivity and the degradation of semantic reasoning abilities. Currently, Speech LLMs often struggle to maintain robustness in the face of nuanced or ambiguous instructions. This sensitivity to instruction variability affects their ability to consistently generate accurate and contextually appropriate output. Additionally, as Speech LLMs are applied to increasingly complex tasks, their reasoning capabilities tend to degrade, particularly in understanding subtle, multi-step contexts. Moreover, Speech LLMs still have a long way to go in achieving true multitask generalizability, where a model can handle tasks it hasn't been explicitly trained on. At present, models often fail to perform certain tasks unless they have been specifically trained for them. Researchers are expected to focus on improving models’ adaptability to diverse instruction formats, enhancing their reasoning capabilities, and enabling more seamless multitask learning across tasks the model has not been trained on.

Another critical avenue for future work lies in extracting richer and finer acoustic information from speech. Although current models can process speech to some extent, their ability to capture the full range of acoustic cues, such as speaker emotion, intonation, and environmental noise, remains limited. Exploring methods to improve the extraction of acoustic information and incorporating these features more effectively into the reasoning process will be key to enabling more nuanced and context-aware speech understanding.

In addition, preference alignment, especially approaches based on reinforcement learning, is a promising direction that has so far received limited attention in Speech LLMs. While it has been widely adopted in the broader LLMs community to align model behavior with human intent, its application in Speech LLMs remains relatively limited. As these models are increasingly deployed in interactive real-world scenarios, it becomes crucial to ensure that their outputs align with user expectations, task-specific requirements, and contextual nuances. RLHF offers a viable mechanism for fine-tuning models based on human preferences and interaction signals, enabling more personalized and context-aware responses~\cite{schulman2017proximal, huang2023survey}. Advancing research in this area could significantly enhance the reliability, controllability, and user satisfaction of Speech LLMs in practical applications.

Together, these areas represent some of the most important directions for future exploration in the development of Speech LLMs. By addressing these challenges, researchers can improve the utility, adaptability, and overall performance of speech-based models, paving the way for more advanced systems capable of deeper understanding and more effective interaction with users.

\section{Conclusion}
\label{sec:conclusion}
This paper presents the first comprehensive survey of Speech LLMs from the perspective of \emph{ speech understanding}. This perspective has been largely overlooked in prior work, which has focused predominantly on speech generation or general-purpose multimodal models. To ground this perspective, we propose the first systematic conceptualization of speech understanding, along with a task-oriented taxonomy that spans informational, functional, and format dimensions. Building on this conceptual foundation, we trace the evolution of modeling approaches, highlighting the transition from early modular and cascaded pipelines to E2E specialized systems, and more recently, to LLM-centric architectures that aim to unify diverse speech understanding tasks. To clarify the current design space, we synthesize recent advances in model architectures, training strategies, and datasets, and further provide an empirical perspective by analyzing how well existing Speech LLMs generalize across a range of speech understanding tasks. Through our own experimental investigation, we identify two pressing challenges: sensitivity to instruction and degradation of semantic reasoning ability of current systems. Both limit the robustness and flexibility of Speech LLMs in real-world applications. By summarizing the field systematically, identifying key challenges, and proposing actionable directions, we hope that this survey can serve as both a conceptual foundation and a practical reference for advancing the development of Speech LLMs toward more general, adaptable, and human-aligned speech understanding systems.

\vfill\pagebreak

% References should be produced using the bibtex program from suitable
% BiBTeX files (here: strings, refs, manuals). The IEEEbib.bst bibliography
% style file from IEEE produces unsorted bibliography list.
% -------------------------------------------------------------------------
\bibliographystyle{IEEEtran}
\bibliography{strings,refs}
\newpage
\section{Appendix}

\subsection{Instruction variants for instruction-sensitivity evaluation}
\label{sec: instruction variants}

For the instruction-sensitivity evaluation in Section~\ref{sec: instruction_sensitivity}, we devise seven instruction prompt variants for each task: the default prompt, a punctuation-altered version, three levels of semantic complexity (simple, neutral, and complex), and two case transformations (lowercase and uppercase). The full content of every variant appears in Table~\ref{tab:instruction_contents}.

\begin{table*}[ht]

\caption{The content of instruction prompt variants for instruction-sensitivity evaluation.}
\label{tab:instruction_contents}
\centering
\scriptsize
\setlength{\tabcolsep}{2pt}           % 列间距
\renewcommand{\arraystretch}{1.05}    % 行高

% 每列宽度：变体列 0.12\textwidth；4 个任务各 0.22\textwidth
\begin{tabular}{p{0.12\textwidth} | p{0.22\textwidth} p{0.22\textwidth} p{0.22\textwidth} p{0.22\textwidth}}
\hline
\textbf{Instruction Variation} &
\textbf{ASR} &
\textbf{S2TT (en→zh)} &
\textbf{SER} &
\textbf{GR} \\
\hline
Default &
Recognize the speech, only output the transcription: &
Translate the speech into Mandarin, only output the translated text: &
Recognize the emotion, in one word: Happy, Sad, Angry or Neutral: &
Recognize the speaker's gender, in one word: Male or Female: \\
\hline
Punctuation &
Recognize the speech, only output the transcription. &
Translate the speech into Mandarin, only output the translated text. &
Recognize the emotion, in one word: Happy, Sad, Angry or Neutral. &
Recognize the speaker's gender, in one word: Male or Female. \\
\hline
Semantic (simple) &
Do asr, only output the transcription: &
Do S2TT to zh, only output the translated text: &
Do SER, in one word: Happy, Sad, Angry or Neutral: &
Do gender recognition, in one word: Male or Female: \\
\hline
Semantic (neutral) &
Transcribe the speech audio, only output the transcription: &
Convert the spoken English into Mandarin Chinese, only output the translated text: &
Identify the speech emotion, in one word: Happy, Sad, Angry or Neutral: &
Identify the speaker's gender, in one word: Male or Female: \\
\hline
Semantic (complex) &
Accurately transcribe every word spoken in the supplied recording, only output the transcription: &
Produce a fluent Mandarin Chinese translation of the provided speech recording in English, only output the translated text: &
Accurately detect the emotional state conveyed in the provided speech recording, in one word: Happy, Sad, Angry or Neutral: &
Accurately detect the gender of the speaker in the provided speech recording, in one word: Male or Female: \\
\hline
Case (lower) &
recoginize the speech, only output the transcription: &
translate the speech into mandarin, only output the translated text: &
recognize the emotion, in one word: Happy, Sad, Angry or Neutral: &
recognize the speaker's gender, in one word: Male or Female: \\
\hline
Case (upper) &
RECOGNIZE THE SPEECH, ONLY OUTPUT THE TRANSCRIPTION: &
TRANSLATE THE SPEECH INTO MANDARIN, ONLY OUTPUT THE TRANSLATED TEXT: &
RECOGNIZE THE EMOTION, IN ONE WORD: Happy, Sad, Angry or Neutral: &
RECOGNIZE THE SPEAKER'S GENDER, IN ONE WORD: Male or Female: \\
\hline
\end{tabular}
\end{table*}

\subsection{Prompts Used for Challenge B: Semantic Understanding Degradation}
\label{sec:challenge2-prompt}
To ensure a fair comparison between Speech LLMs and their base text LLMs, we adopt a two-stream evaluation setup in which both models receive matched semantic inputs under their respective modalities. The Speech LLMs receive audio inputs, while the base LLMs are provided with the corresponding ASR transcripts or ground truth transcriptions. The same task instruction is used in both streams.

We use different prompt templates depending on the model, based on the official prompt lists reported for each model.

\subsubsection{SALMONN (Vicuna)}
\begin{itemize}
    \item ASR Task with Speech Input:
    \begin{quote}
        \texttt{Please transcribe the speech into a written format.}
    \end{quote}
\end{itemize}
\paragraph{Speech Translation (ST)}
We adopt the following task prompts:

\begin{itemize}
    \item Translation Task with Speech Input :
    \begin{quote}
        \texttt{Listen to the speech and translate it into Chinese.}
    \end{quote}

    \item Translation Task with GT/ASR Text Input:
    \begin{quote}
        \texttt{Translate the text into Chinese.}
    \end{quote}
\end{itemize}

\paragraph{Spoken Language Understanding (SLU)}
For SLU, we use the official prompt template from the MMSU benchmark:

\begin{quote}
\texttt{
Choose the most suitable answer from options A, B, C, and D to respond to the question in the next line. \\
You should only choose A or B or C or D. Do not provide any additional explanations or content.}

\vspace{0.3em}
\texttt{
Question: \{question\}\\
A. \{choice\_a\} \\
B. \{choice\_b\} \\
C. \{choice\_c\} \\
D. \{choice\_d\}
}
\end{quote}

During evaluation, the model's final answer is extracted by identifying the first capital letter (A, B, C, or D) mentioned in its output.

\subsubsection{Qwen2.5-Omni (Qwen2.5-Instruct)}
We adopt the officially released prompt templates (with systems input) for Qwen2.5-Omni and its base variant Qwen2.5-Instruct, as reported in its model card and demos. These prompts are tailored for different modalities and tasks.

\begin{itemize}
    \item ASR Task with Speech Input:
    \begin{quote}
    \texttt{Recognize the speech, only output the transcription, do not ask any other things. Your answer should appear between <ASR\_TRANS></ASR\_TRANS>. Example: Input audio "I'm Qwen." Answer: <ASR\_TRANS>I'm Qwen.</ASR\_TRANS>\textbackslash n Now please recognize the input speech.}
    \end{quote}
\end{itemize}
\paragraph{Speech Translation (ST)}
\begin{itemize}
    \item Translation Task with Speech Input:
    \begin{quote}
    \texttt{Directly translate the speech to Chinese transcription, only output the translation, do not ask any other things. Your answer should appear between <translate\_to\_cn>. Example: Input audio: "I'm Qwen". Answer: <translate\_to\_cn>\begin{CJK}{UTF8}{gbsn} {我是}\end{CJK}Qwen</translate\_to\_cn>.\textbackslash n Now please recognize the input speech.}
    \end{quote}
    \item Translation Task with GT/ASR Text Input:
    \begin{quote}
        \texttt{Directly translate the text to Chinese, only output the translation, do not ask any other things. Your answer should appear between <translate\_to\_cn>. Example: Input text "I'm Qwen." Answer: <translate\_to\_cn> \begin{CJK}{UTF8}{gbsn} {我是}\end{CJK} Qwen\\</translate\_to\_cn>\textbackslash Now please translate the input text:}
    \end{quote}
\end{itemize}

\paragraph{Spoken Language Understanding (SLU)}
\begin{quote}
\texttt{
You are given a question with multiple choices. Choose the single correct choice (A/B/C/D) that answers the question best. ONLY output the letter (A/B/C/D), no explanation.\\
Example: Input question "Which city is the capital of France?" Choices: A. Berlin B. Madrid C. Paris D. Rome. Answer: C\\
}
\texttt{
Now please answer:\\
Question: \{question\}\\
A. \{choice\_a\} \\
B. \{choice\_b\} \\
C. \{choice\_c\} \\
D. \{choice\_d\}
}
\end{quote}

As with SALMONN, during evaluation, the model's final answer is extracted by identifying the first capital letter (A, B, C, or D) mentioned in its output.

\end{document}